\documentclass[fleqn,10pt]{wlscirep}
\usepackage[utf8]{inputenc}
\usepackage[T1]{fontenc}
\usepackage{bbm}
\newcommand{\ind}{\mathbbm{1}}

\title{Causal Impacts of Protected Bike Lanes on Cycling Behavior with Demographic Disparities}

\author[1,*]{Marcel Moran}
\author[1]{Malik Salman}
\author[1,2,*]{Takahiro Yabe}
\affil[1]{Center for Urban Science and Progress, Tandon School of Engineering, New York University, Brooklyn, NY, 11201, USA}
\affil[2]{Department of Technology Management and Innovation, Tandon School of Engineering, New York University, Brooklyn, NY, 11201, USA}

\affil[*]{marcel.moran@nyu.edu and takahiroyabe@nyu.edu}

\keywords{Cycling, Causal Inference, Planning}

\begin{abstract}
Cities around the world face significant barriers to grow urban cycling, including competing budgetary priorities and car-centric streets. Thus, when making decisions regarding the installation of bicycle infrastructure, it is crucial to understand if and to what extent different bicycle-lane types increase bicycle ridership. However, associations between bicycle infrastructure and bicycle ridership have primarily been studied in the context of individual lanes and corridors, or when analyzed at the scale of entire cities, generalized across different bike-lane types. Drawing upon 72 million bikeshare trips from Citi Bike in New York, we demonstrate that there is an approximately 18\% increase in bikeshare trips at adjacent stations in the 12 months following the installation of protected bike lanes (those with a physical barrier between cyclists and automobile traffic) and a 14\% increase associated with painted bike lanes (where a line of pavement marking is present) and `sharrows' (where a normal traffic lane is marked with a bike stencil). However, using a difference-in-differences analysis, we detect a causal effect on bikeshare ridership only following the installation of protected bike lanes, with an average monthly increase of 379 rides per station ($p<0.001$). Despite this causal effect being pronounced among census block groups with higher percentages of older adults (688 rides per month per station, $p<0.001$), the causal effect of protected bike lanes on bikeshare ridership is absent in census block groups where the percentage of Black residents is medium to high. Taken together, these findings indicate that planners must emphasize protected bike lanes to spur ridership, and simultaneously target policies and programming to communities of color, to ensure that such infrastructure makes urban cycling a viable option for all residents.  

\end{abstract}
\begin{document}

\flushbottom
\maketitle

\thispagestyle{empty}

\section*{Introduction}

Increasing urban cycling can benefit cities by reducing air and noise pollution, carbon emissions, and roadway fatalities, relative to automobiles.\cite{pucher2017cycling} This is of particular importance given the transportation sector now represents the largest share of carbon emissions in the U.S.,\cite{EPA2024} and has been stubborn in terms of decarbonization.\cite{CBO2022transport} In addition, cycling and pedestrian fatalities have remained high in the U.S. relative to peer nations,\cite{buehler2017trends, schneider2020united} as have deficits in daily physical activity.\cite{kapteyn2018they} Despite this constellation of motivations for increasing urban cycling, it has remained low in American cities, relative to European counterparts.\cite{buehler2012international} This is particularly the case in terms of race and ethnicity; people of color are under-represented among cyclists nationally.\cite{buehler2012international, sadeghvaziri2024active} Part of this disparity is potentially tied to infrastructure; the primary means of both safeguarding existing cyclists and encouraging new cyclists is by providing bike lanes, which designate semi- or fully-exclusive space for bicycle riding within the public right of way.\cite{pucher2008making} Though some U.S. cities have built large bike-lane networks (including Boston, Washington, D.C., Portland, OR, and San Francisco), many cities still provide little in the way of physical infrastructure for cycling. Further, a study of the provision of bike lanes in 22 large American cities determined that census block groups with higher percentages of non-white residents had lower access to bike lanes.\cite{braun2019social} There is also growing evidence that protected bike lanes specifically can lead to increased ridership diversity. A mailed and intercept survey in Boston found a strong preference for protected bike lanes among non-white respondents,\cite{lusk2017biking} and video recording and intercept survey of riders on bike lanes in Brooklyn indicated high non-white and local ridership.\cite{noyes2014cycling} Cycling in the U.S. also remains overrepresented by young people.\cite{Garrard2021} Though there has been fewer quantitative studies on bicycle infrastructure and older adults, cities that have implemented a broad range of programs to encourage bicycling, including protected bike lanes, have seen the share of trips by older adults increase.\cite{pucher2010infrastructure} Indeed, this area of bicycle research requires further scrutiny, particularly as America's population ages in the coming decades.\cite{ortman2014aging} 

Active-transportation research provides consistent evidence of a positive association between bike lanes and bike ridership. At the scale of a single bike lane, a difference-in-differences analysis found that a new bike lane in Boston causally increased bikeshare ridership by 80 percent.\cite{karpinski2021estimating} In New York City, for each additional mile of bike lanes installed in Manhattan, 285 more bikeshare trips were generated.\cite{xu2020longitudinal} Further, analyzing bicycle traffic at newly-installed protected lanes across five U.S. cities detected increases in cycling ranging from +21 percent to +171 percent.\cite{monsere2014lessons} In addition, in the wake of the COVID-19 pandemic, data from bicycle counters across 106 European cities indicated that the roll-out of provisional or `pop-up' bike lanes\cite{buehler2021covid, moran2022treating} was associated with increases in cycling between 11-48\%.\cite{kraus2021provisional} 

There are three important features of this existing literature that inform the design of this study. First, not all analyses differentiate between bike-lane type, at times treating painted and protected bike lanes as equivalent. Second, the variation in spatial scale (e.g. block, corridor, etc.) can make it difficult to compare findings across case studies, or draw larger insights about how bike-lane type influences behavior at the scale of an entire city. Third, attempts at linking bikeshare trips to bike lanes, including difference-in-differences analyses,\cite{chahine2025effect} have yet to incorporate broader sociodemographics of the areas where trips occur. For these reasons, we sought to probe the relationship between bike lanes and bikeshare ridership at the city scale, while maintaining block-level specificity as to bike-lane type, and incorporating a range of likely-relevant sociodemographics. Based on works cited above, we hypothesize that not only will the expansion of bike lanes in New York City be associated with increases in bikeshare trips generally, but that there will be detectable spatial differences in such increases based on what type of bike lane each bikeshare station lies adjacent to. Indeed, we hypothesize that bikeshare trips which begin adjacent to \textit{protected} bike lanes will increase to a greater extent than those which begin adjacent to either painted bike lanes or sharrows. Further, given existing racial disparities in biking, we hypothesize that in areas in which non-white persons make up a larger percentage of residents, the effect of protected bike lanes on bikeshare ridership will be less pronounced. 

With this context, we pursue a simultaneously comprehensive and fine-grain analysis of the relationship between bike-lane types and bikeshare trips, in the case of New York City, which maintains both a massive and expanding bike-lane network (see \textbf{Figure 1A}), as well as North America’s largest bikeshare system (Citi Bike). Specifically, we analyze roughly 72 million publicly-available Citi Bike trips, in conjunction with publicly-available bike-network data, in order to detect how trips change over time based on what type of bike lanes they begin adjacent to. Given cycling’s ability to increase urban transportation's safety, sustainability, and physical health, if and to what extent different types of bike lanes are associated with increases in cycling is paramount to assessing their value, and thus their prioritization by transportation planners. Indeed, bike-lane types require different funding levels and more or less-extensive design and construction processes, which means that evidence as to which type of bike infrastructure is most associated with increased cycling is critical amid constrained budgets and competing street demands (e.g. bus lanes, automobile parking, loading zones, etc.).  

The following methods generate three core findings. First, correlational analyses demonstrate that following bike-lane installation, there are increases in bikeshare ridership at stations adjacent to protected bike lanes (+18\%), and painted bike lanes and sharrows (+14\%). However, a difference-in-differences (DiD) analysis demonstrates that only \textit{protected} bike lanes exhibit a causal effect on bikeshare ridership. Third, when testing for heterogeneous effects, protected bike lanes only demonstrate a causal effect on bikeshare ridership in the the areas with the lowest percentage of Black residents, and simultaneously is more pronounced in the areas with the highest percentage of older adults (60-79). Collectively, this suggests that cities should both emphasize protected bike lanes over painted bike lanes and sharrows to increase bike ridership, and simultaneously consider what range of geographically-targeted policies and programs can ensure that this infrastructure leads to growth in cycling for all citizens, especially communities of color. 

\section*{Results}

In New York, bike lanes are primarily provided as either protected bike lanes (those with some sort of physical barrier and/or buffer from automobiles), painted bike lanes (those lying adjacent to automobile traffic and separated only by a painted line), or ‘sharrows’ (normal traffic lanes marked with bicycle stencils) (see \textbf{Figure 1B}). The analysis begins by spatially overlaying all Citi Bike stations on top of New York City’s bike-lane network (both as of the close of 2024). Next, all bike lanes within 100 meters of each Citi Bike station, along the street grid, were identified. Stations were then categorized by the type of bike lane within this distance threshold (protected, painted, sharrow, or none), and if multiple types of bike lanes were present within 100 meters, the station was categorized based on the highest-quality bike lane (i.e. protected first, painted second, and sharrow third). For each of those Citi Bike stations, now categorized by their adjacent bike-lane type, all trip starts were downloaded between the years of 2013 and 2024. More details on the assignment of bike lanes (treatment variable) to bike stations can be found in Supplementary Note 2.1.  

\begin{figure}[t]
    \centering
    \includegraphics[width=.95\linewidth]{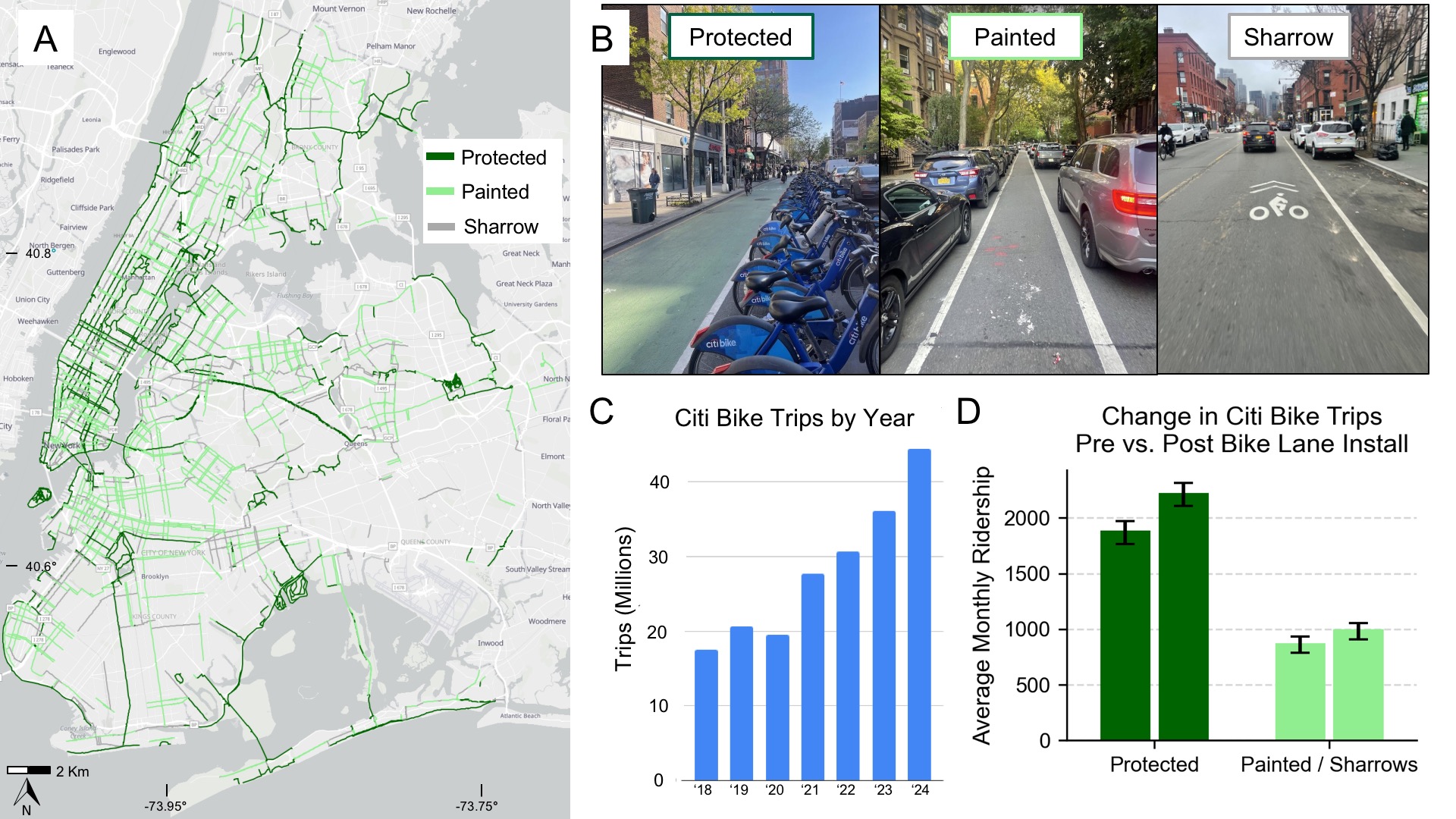}
    \caption{A. Map of the spatial distribution of bike lane types in New York City, including protected bike lanes (where a physical barrier is provided between cyclists and automobile traffic), painted bike lanes (where a line of paint designates space for cyclists adjacent to automobile traffic), and sharrows (where a bicycle stencil is marked on the pavement of a shared traffic lane). B. Street-level photographs of the three bike lane types in New York City: Protected, Painted, and Sharrows (all photos taken by the authors). C. Growth in annual Citi Bike trips, 2018-2024. D. Change in average Citi Bike trips, from the 12 months prior to the 12 months after bike-lane installation, by the type of bike lane adjacent to each bikeshare station. There is a positive association between the installation of protected bike lanes, as well as painted bike lanes or sharrows, with Citi Bike trips, of approximately 18\% and 14\% respectively.}
    \label{fig:enter-label}
\end{figure}

Taken together, these data-processing steps allow for all Citi Bike stations to be analyzed both in terms of what type of bike lane they lie adjacent to, as well as trip counts for each month across this time period. In general, these years included both consistent growth in Citi Bike trips (hitting a system record of roughly 45 million trips in 2024, see \textbf{Figure 1C}), and steady expansion of bike lanes throughout New York City, concentrated in Manhattan, but also including significant growth in western Brooklyn, western Queens, and the South Bronx (see \textbf{Figure 1A}). Because Citi Bike is not currently present on Staten Island, that borough is excluded from this analysis.  

\subsection*{Bike Lane Installation is Correlated with Bikeshare Trip Growth}
New York City’s dataset for its bike-lane network includes the month and year in which each bike-lane segment was installed. That attribute allows for a comparison of Citi Bike trip counts before and after bike lanes of the three different types were installed. Thus, for each Citi Bike station, we calculated the difference between trip counts for the 12 months prior to a bike lane adjacent to it being installed, to the 12 months afterwards. 

In the aggregate, we find that Citi Bike stations adjacent to protected bike lanes exhibited the largest average difference in the 12 months after installation (+340.6 rides per month), followed by stations adjacent to painted bike lanes and sharrows (+122.3 rides per month) (see \textbf{Figure 1D}). This computes to gains of roughly (+18\%) and (14\%) respectively. This correlational analysis suggests that both protected and painted bike lanes are associated with increases in ridership at nearby stations. However, this analysis could be confounded by various spatial and sociodemographic factors. For example, bike lanes may be selectively placed in areas where there is already large and growing ridership. To rigorously analyze the effects of bike lanes on ridership, causal analysis is required. 

\subsection*{Difference-in-Differences Analysis Demonstrates Causal Effect of Trip Growth from Protected Lanes Only}
To further scrutinize the relationship between bike-lane type and Citi Bike trips, we conducted a causal analysis that consists of propensity matching and difference-in-differences (DiD) analysis. To match treated bike stations with bike lanes installed nearby with similar bike stations without bike lanes nearby, we used propensity score matching.\cite{caliendo2008some} Propensity scores were estimated using the following covariates: total population, population density, median household income, percent Black, percent Hispanic, percent over the age of 25, percent with a bachelor's degree, number of Citi Bike stations within 200 meters, and number of dining establishments within 200 meters. The number of stations within 200 meters ($\beta = -0.73$, $p<0.001$) and the number of dining options within 200 meters ($\beta = 0.0035$, $p<0.001$) were found to be predictive of bike lane installment. Details on the formulation and logistic regression results for the propensity score estimation can be found in Supplementary Note 2.2 and Supplementary Table S1. 

Propensity score matching resulted in 234 pairs of Citi Bike stations, ensuring that each were alike in terms of a range of attributes (at the level of a Census Block Group). We then further refined this set of station pairs by selecting only those in which one of the stations includes an adjacent bike lane (either protected, painted, or sharrow), and the other lying adjacent to no bike lane at all. We also excluded station pairs if there were stretches of at least three months without any trips at all, to avoid stations disrupted by construction. That sifting resulted in 216 total station pairs for the DiD analysis. To validate the quality of the match, we assessed covariate balance between treated and control stations using standardized mean differences (SMDs), confirming that all matched covariates fell below conventional thresholds for imbalance (SMD < 0.1), as shown in Supplementary Figure S1. Furthermore, the near-zero difference in propensity scores between the treatment and control group distributions shown in Supplementary Figure S2 further supports the effectiveness of our matching approach.

To estimate the causal effects of protected and painted bike lanes on changes in ridership, we use a two-way fixed effects difference-in-differences model with time- and unit-fixed effects:
\begin{equation}
Y_{it} = \alpha + \beta_0 W_{it} \cdot \ind(C_i = 0)
 + \beta_1 W_{it} \cdot \ind(C_i = 1)+ u_i + \omega_t + \epsilon_{it}
\end{equation}
where $Y_{it}$ represents the number of rides starting from station $i$ in month $t$. $W_{it}$ is a binary dummy variable that is 1 if station $i$ is treated and the installation year and month have been reached. 
The variable $C_i \in \{0,1\}$ denotes the type of bike lane treatment assigned to station $i$, with $C_i = 0$ indicating a painted bike lane and $C_i = 1$ indicating a protected bike lane. The coefficient $\beta_0$ thus captures the average treatment effect of painted bike lanes on ridership, while $\beta_1$ captures the effect of protected bike lanes. 
$u_i$ accounts for the station-level fixed effects, and $\omega_t$ represents time fixed effects (year and month). Finally, $\epsilon_{i,t}$ represents the error term. 
By allowing separate coefficients for the two categories, the model accommodates heterogeneous treatment effects by infrastructure type. Supplementary Note 2.3 provides more details on the methods.

This formulation takes advantage of each bike lane's install month and year to compare Citi Bike trips from the pre-installation period to the post-installation period, comparing the paired stations (one with a bike lane, one without). The results of the DiD analysis, shown in \textbf{Figure 2C} demonstrates that growth in Citi Bike trip counts were statistically significant for the stations which lie adjacent to protected bike lanes with an average treatment effect ($\beta = 379.88$, $p<0.001$), compared to the paired stations without bike lanes, as shown in \textbf{Figure 2C}. Moreover, trip growth following the installation of painted bike lanes and sharrows, relative to paired stations without bike lanes, was not statistically significant ($\beta = -141.18$, $p=0.156$), demonstrating no benefit over stations without any bike lanes. Detailed regression results are shown in Supplementary Table S2. 

To assess the robustness of our DiD estimates, we conducted a placebo test by randomly shuffling the installation dates of the bike lanes while keeping all other data fixed. For each of the 500 iterations, we reassigned treatment timing at random and re-estimated the DiD model using the same specification as equation (1). This placebo-based threshold allows us to evaluate whether the observed treatment effect is likely to have arisen by chance due to time-varying confounders or unobserved shocks. The results in \textbf{Figure 2C} show that the effect is insignificant, supporting the claim that our results are not confounded by unobserved variables. Detailed description of the placebo tests and results are presented in Supplementary Note 2.4 and Figures S3 and S4.

\begin{figure}
    \centering
    \includegraphics[width=.95\linewidth]{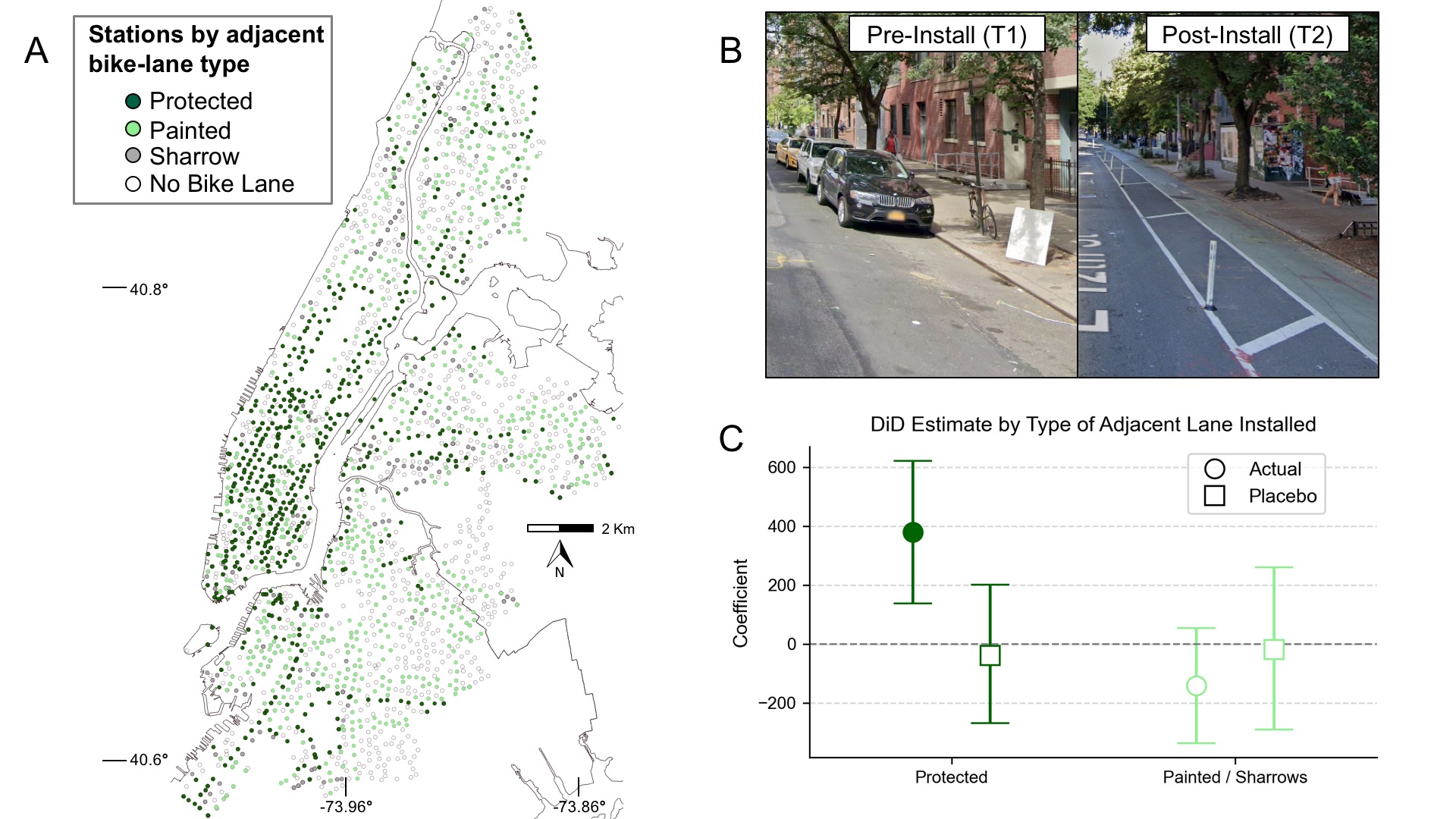}
    \caption{A. Citi Bike Stations as of the close of 2024, colored by the bike-lane type they lie adjacent to (protected, painted, sharrow, or none). B. Photographs of an example street in New York City both before and and after protected bike-lane installation (photos drawn from Google Maps). C. Difference-in-Differences (DiD) analysis of the change in bikeshare trips following the installation of protected bike lanes (left) and painted bike lanes and sharrows (right). Propensity score matching was used to pair bikeshare stations which are similar in a number of regards, but differ in terms of the presence or absence of a recently-installed, adjacent bike lane. The coefficients demonstrate that there is a causal effect only for protected bike lanes on ridership, one not detected for painted bike lanes and sharrows.}
    \label{fig:enter-label}
\end{figure}

\subsection*{Growth in Bikeshare Trips Exhibits Demographic Disparities}
Only the 128 Citi Bike stations treated with protected lanes and their matches were used for this part of the analysis. For each station, a range of sociodemographic information from the US Census and American Community Survey were downloaded at the level of Census Block Group (CBG). This allows analysis of such variables in the context of the DiD; specifically, determining whether the causal effect of protected bike lanes on Citi Bike trips holds across thresholds of different variables. We use the following DiD specification:
\begin{equation}
Y_{it} = \alpha + \sum_{{b}} \beta_{b} W_{it} \cdot \ind(b_i = b) + u_i + \omega_t + \epsilon_{it}
\end{equation}
where $Y_{it}$ is the number of rides starting from station $i$ in month $t$, and $W_{it}$ is a binary indicator for whether a protected bike lane has been installed for station $i$ as of month $t$. The variable $b_i$ denotes the bin assignment for station $i$ based on the value of the moderating variable, categorized into three levels: low, medium, and high. The term $\mathbf{1}\{b_i = b\}$ is an indicator function equal to 1 if station $i$ falls into bin $b$, and 0 otherwise. The coefficient $\beta_b$ therefore represents the average treatment effect of protected bike lanes for stations in bin $b$.

This heterogeneous effect analysis demonstrated that the causal effect of protected bike lanes on bikeshare trip growth is strongest when the percentage of Black residents is lowest ($\beta = 464.00$, $p<0.001$), as shown in \textbf{Figure 3}. However, the placebo test rejected the significance of the causal effect of protected bike lanes based on CBG percentages of Hispanics, or percentage with a bachelors degree, or household income (details in Supplementary Note 3), despite the strong and positive correlation between percent Black residents and percent Hispanic residents, and a negative correlation between percent Black residents and percent with a bachelors degree, and household income (Supplementary Figure S5). This result emphasizes the specific relationship between communities with high proportion of Black residents to bike lanes and bikeshare ridership. 

An additional variable tested for heterogeneous effects is the percent of older adults (ages 60-79), given there is evidence that older adults may be less likely to bike unless and until improved bicycle lanes are provided. This variable had a non-significant correlation with the other sociodemographic variables that were tested in the previous paragraph (including race, income, and Bachelor degree). Using propensity score matching and the heterogeneous DiD analysis, we found that census block groups with a high percentage of older adults saw significant positive effects of the protected bike lane installment ($\beta = 684.19$, $p<0.001$), as shown in \textbf{Figure 4}. Placebo tests confirm the robustness of this result. Supplementary Tables S3 to S7 and Figures S6 to S15 show the details of the heterogeneous causal inference results. 

\begin{figure}
    \centering
    \includegraphics[width=.9\linewidth]{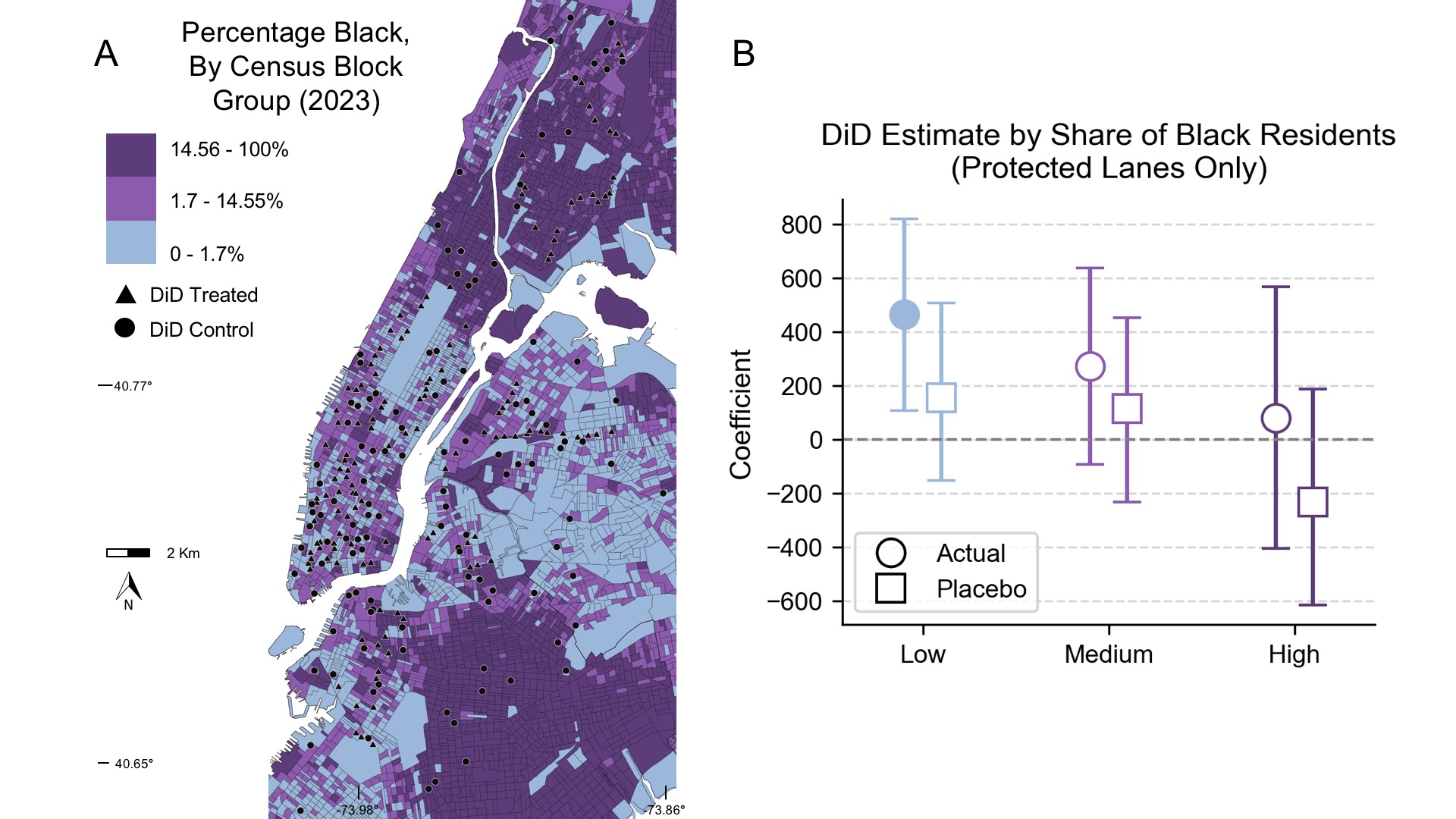}
    \caption{A. Map of the percentage of Black residents by Census Block Group in New York City (2023), overlaid with Citi Bike stations serving as treated (triangles) and controls (circles) for the DiD analyses. B. Heterogeneous Effects of protected bike lane installation on bikeshare ridership based on percentage of Black residents in each Census Block Group. The coefficients demonstrate that the causal effect of protected bike lanes on bikeshare ridership is only present in Census Block Groups with the lowest share of Black residents.}
    \label{fig:enter-label}
\end{figure}

\begin{figure}
    \centering
    \includegraphics[width=.9\linewidth]{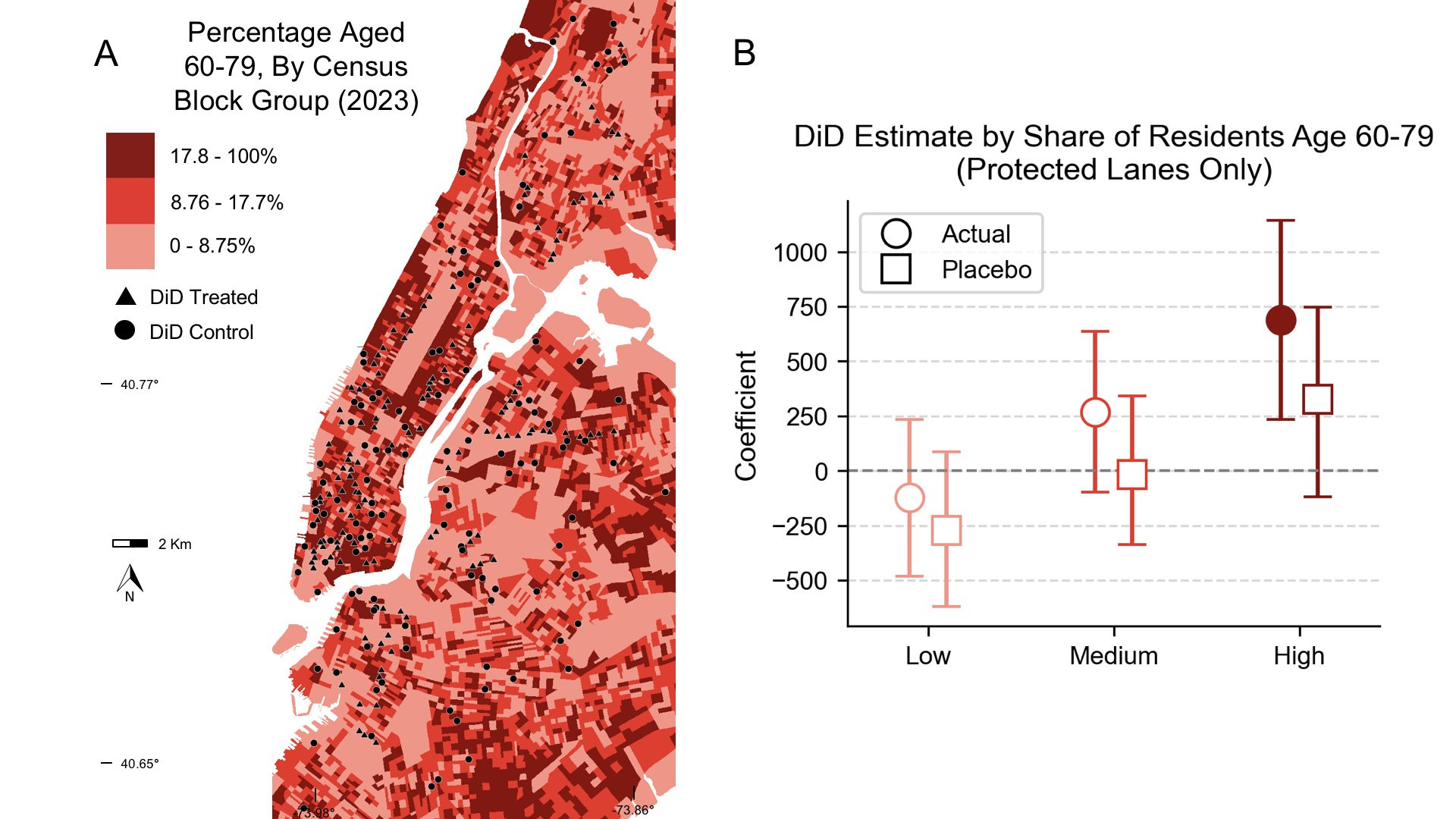}
    \caption{A. Map of the percentage of older adults (60-79) by Census Block Group in New York City (2023), overlaid with Citi Bike stations serving as treated (triangles) and controls (circles) for the DiD analyses. B. Heterogeneous effects of protected bike lane installation on bikeshare ridership based on percentage of older adults (60-79). The coefficients demonstrate that the causal effect of protected bike lanes on bikeshare ridership is more pronounced in Census Block Groups in which the percentage of older adults is the highest.}
    \label{fig:enter-label}
\end{figure}

\section*{Discussion}
Analyzing roughly 72 million trips from Citi Bike data, and linking those trips to the bike lanes adjacent to where each trip starts, we demonstrate at a municipal scale and with block-level granularity that protected bike lanes causally increase bikeshare trips, an effect not present for painted bike lanes and sharrows. Further, by incorporating sociodemographic variables from the Census Block Groups where each Citi Bike station is located, we show the causal effect of protected bike lanes on bikeshare trips is present only where the percent of Black residents is the lowest, and more pronounced where the percent of older adults (60-79) is highest. These findings leverage data from the largest bikeshare system in North America to illustrate the specific relevance of protected bike lanes, compared to lower cost and less design-intensive alternatives. Indeed, in the context of transportation planning, budgetary and political calculations often entail installing painted bike lanes and sharrows over protected bike lanes, though this study emphasizes the achieved benefits of such facilities may be far less meaningful.

 Our analysis further demonstrates that the pro-cycling effect of protected bike infrastructure is not distributed uniformly across New York City's racial geography. Indeed, heterogeneous effects of protected bike lanes on bikeshare ridership were not detected in relation to the percentage of Hispanic residents, the percentage of residents with bachelors degrees, or based on household income, which suggests a specific challenge of expanding cycling within African-American communities. In this vein, existing scholarship  consistently documents the ways in which bicycle planning in the United States does not adequately incorporate the preferences of minority neighborhoods,\cite{hoffmann2016bike} as well as how discriminatory policing of bicyclists of color deters more diverse bike ridership.\cite{lugo2018bicycle, lubitow2019sustainable, barajas2021biking} In response, planners must work to ensure bikeshare (and bicycling generally) is an attractive travel option for all residents. This includes bikeshare membership discounts (which Citi Bike does offer), bicycle education and training programs in local schools and community centers, promotional events, secure bike parking (a current deficit in New York), and refurbished bicycles available for purchase at affordable prices.\cite{merce2023training, mcneil2017breaking} Further, the planning process for new bike lanes must ensure voices throughout different communities are heard, such that the resulting bike-lane network can draw large and diverse ridership. 
 
 In regards to age, this study also points to the crucial role that bike infrastructure can play in increasing the number of older adults who bike. In Northern European cities such as Amsterdam, Copenhagen, and Hamburg, meaningful percentages of older adults bicycle, in large part owing to expansive and high-quality bike lanes.\cite{pucher2010infrastructure} The pronounced effect of protected bike lanes on bikeshare ridership in the CBGs with the highest share of older adults detected here may indicate the infrastructure's encouraging effect on riders highly sensitive to safety concerns.\cite{ottoni2021safety, van2018environmental}  

Future research can take a number of directions to build on these analyses. First this approach can be applied to wherever bike-lane and bikeshare data are publicly available, including for those cities for which Lyft also manages large bikeshare systems (including Chicago, San Francisco, Boston, and Washington, D.C.). In addition, the increased rate of bikeshare \textit{electrification} should also be examined in terms of its effects on ridership; existing studies indicate e-bikes can appreciably increase ridership, particularly among older adults.\cite{van2019bikes} Furthermore, available datasets on bike lanes and bikeshare create exciting opportunities to consider how infrastructure and mobility services are differentially provided within neighborhoods, including those underserved by transit, or with low rates of car ownership.\cite{firth2021were} There is also the chance to further link infrastructure and ridership to safety outcomes, given prior evidence of widespread bike lanes conferring citywide decreases in bicycle-involved collisions.\cite{marshall2019cities} Research partnerships, such as with the operator Lime, have thus far similarly provided evidence for the benefit of protected bike lanes on dockless bikeshare.\cite{lime2024} In addition, bike share is of course only one category of cycling; data collected from other bicyclists, including via mobile applications like Strava,\cite{garber2023bicycle} can further delineate the effect of bicycle infrastructure on ridership at the municipal scale.

There are several limitations to this analysis. The first is the assumption that if a Citi Bike station lies adjacent to a specific type of bike lane, that reflects the infrastructure of the entire trip in question. Second, the three categories used here for bike lanes in New York City — protected, painted, and sharrows — inherently mask some heterogeneity in these facilities. For example, some bike lanes are `protected' by plastic posts, others by a lane of parked cars, and others by concrete barriers, each of which likely influence cycling behavior to different extents. Third, propensity score matching to develop station pairs for the DiD analyses inherently involves subjectivity — what variables are used to match routes, what variables are omitted, and how attributes are weighted. To address this limitation, we test for pre-treatment parallel trends with placebo tests, and restrict our sample to stations with stable usage patterns prior to treatment, increasing confidence that post-treatment differences reflect causal impacts rather than pre-existing growth trajectories.

\section*{Methods}
\subsection*{Citi Bike Data}
Citi Bike data is available for download for all trips in the system's history, beginning in 2013. Trip data includes start date and time, end date and time, as well as start and end station. A number of other variables either were provided at one point and since discontinued (such as rider gender), or have been subsequently added and are not present in previous years (such as bike electrification status). Across years, station names have remained largely consistent. Filtering Citi Bike trip data by unique station name allows the extent of all stations to be mapped for a given point in time. Citi Bike trip data can be downloaded from the following link: \url{https://citibikenyc.com/system-data}.  

\subsection*{New York City Bike Lanes}
The City of New York publishes a map and dataset of its current bike lane network on its open-data portal, which can be downloaded and analyzed within GIS platforms. Each bike-lane segment includes installation month and year, as well as `facility class,' an attribute which reflects whether the segment is protected, painted, or a sharrow (with some heterogeneity within these classes). All bike lane data for the analyses undertaken here reflect the city's network at the close of 2024. The bike lane network data of New York City can be downloaded from the following link: 
\url{https://data.cityofnewyork.us/dataset/New-York-City-Bike-Routes-Map-/9e2b-mctv}.

\subsection*{Correlational Analysis of Citi Bike Trips and Bike-Lane Types}
After loading both unique Citi Bike Stations and New York's bike lane segments into a GIS platform, a spatial join was conducted such that any bike lane segments within 100 meters (along the street grid) of a Citi Bike station were joined. Taking into account the installation month and year of each bike-lane segment, as well as bike-lane type (combining painted bike lanes and sharrows), that allowed for a comparison of Citi Bike trips for the 12 months prior to installation, to the 12 months after, which were averaged across types.  

\subsection*{DiD Treatment Assignment}
We begin with a dataset of all Citi Bike stations and their corresponding bike lanes (within 100 meters). Since Citi Bike trip data is available for the full years of 2014 to 2024, we make sure that those units classified as treated had nearby bike installation/s after 2014 and before 2024. This leaves at least 12 months before and 12 months after the installation for our difference-in-differences (DiD) analysis. If the station had at least one protected lane installed nearby, it is classified as a treated unit with a protected bike lane. If the station did not have a protected lane installed, but instead saw a painted intervention (this includes both painted lines and sharrows) implemented, the bike station is labeled as a treated unit but with a painted bike lane. If only one installation took place, this installation’s date is used for analysis. If more than one occurred, the earliest installation date of the relevant class is used. 

A similar approach is taken to identify eligible control units. The filter now selects installation dates that fall before 2014, as well as stations that never had bike lanes installed nearby (untreated across time). As before, stations are classified into categories: treated with protected bike lane, treated with painted bike lane, and untreated. Untreated in this context means not treated in or between 2015 and 2023. In the event that a station had a painted lane installed during the treatment period, but was already in the protected lane type before this period, it was considered untreated. And vice versa: a painted lane station upgraded to a protected lane station during the treatment period is considered treated. 

Once treated and control groups were selected, we merged the classified stations with the trip data. We retrieve the first month and year that a station saw ridership and keep only those treated stations where the station came before the bike lane installation. This ensures that there is ridership to compare pre- and post-install. Lastly, we take into account extended station closures (due to construction, maintenance, etc.) by dropping stations that have more than three consecutive zero-ridership months after ridership first appears. We end this section of the methods with 234 treated units and 1,414 eligible control stations.

\subsection*{Propensity Score Matching}
Before performing the formal difference-in-differences (DiD) analysis, we refine the selection of control stations using propensity score matching (PSM) with replacement. This approach allows each control station to be matched with multiple treated stations, increasing the robustness of our analysis. We focus on one-to-one matching with replacement as it generally leads to greater bias reduction than one-to-one matching without replacement. Specifically, we employ nearest-neighbor PSM to match treated stations with control stations that exhibit similar characteristics. The matching criteria are the following: median household income, percent [of the population] that identifies as black, percent that identifies as Hispanic, percent that has attained at least a bachelor’s degree, stations within 200m, dining establishments within 200m, and population density.

In order to perform PSM, we utilize several data sources. First, we leverage 2023 American Community Survey (ACS) data for the state of New York — this provides us with census-block-group (CBG) level data on median household income, total population, black population, hispanic population, and educational attainment. The last of these references the population of people with degrees equal to, or more advanced than, a bachelor's. We use the CBG information from the U.S. Census Bureau's TIGER/Line Shapefiles interface. Using this information, we are able to calculate population density, as well as sociodemographic percentages relative to total population.

Another covariate we account for is the density of POIs. From Foursquare’s POI data, we isolate dining establishments specifically, and then perform a k-d tree query to locate the number of dining establishments within 200 meters of each station. We use a similar approach to calculate network density: using a k-d tree model, we identify the number of Citi Bike stations within 200 meters of each station. Lastly, we exclude those CBGs where population is equal to zero and impute median household income for those CBGs for which it is absent from the dataset (this occurs 251 times, but never for a treated unit). We end this part of the process with 216 treated units and 1,326 eligible control units.

To examine the relationship between our covariates and the likelihood of treatment, we fit a logistic regression model where the treatment variable is regressed on the covariates:

\begin{align}
\log\left(\frac{P(\text{Treated} = 1)}{1 - P(\text{Treated} = 1)}\right) 
&= \beta_0 
+ \sum_{i=1}^{g} \beta_{1,i} \cdot \text{Population Density}_i \nonumber \\
&\quad + \beta_2 \cdot \text{Median Household Income} 
+ \beta_3 \cdot \text{Percent Black} \nonumber \\
&\quad + \beta_4 \cdot \text{Percent Hispanic} 
+ \beta_5 \cdot \text{Dining Within 200m} \nonumber \\
&\quad + \beta_6 \cdot \text{Stations Within 200m}
\end{align}

This model has an accuracy score of 0.859, demonstrating reasonable predictive performance. Detailed regression results are presented in Supplementary Table S1. Once a propensity score is assigned to each Citi Bike station, we turn to nearest neighbor matching to match each treated unit with a control unit. After the matching process, we assess the covariate balance between the treatment and control groups to ensure comparability. The standardized mean differences (SMD) plot in Supplementary Figure S1 and the propensity score distribution shown in Supplementary Figure S2, confirm the achievement of a balanced covariate distribution between the groups.

More specifically, The SMD measures the difference in means in the unit of pooled standard deviation for a specific covariate, and is calculated as follows:

\begin{equation}
\text{SMD} = \frac{\bar{x}_{treatment} - \bar{x}_{control}}{\sqrt{\frac{s_{treatment}^2 + s_{control}^2}{2}}}
\end{equation}
where $\bar{x}_{treatment}$ and $\bar{x}_{control}$ denote the sample mean of the covariate in treated and untreated subjects, respectively, whereas $s_{treatment}^2$ and $s_{control}^2$ denote the sample variance of the covariate in treated and untreated subjects, respectively. A SMD below 0.10 for a given covariate suggests sufficient overlap between the treatment and control groups, indicating that the matching process has successfully minimized bias in the covariate distributions. Furthermore, the near-zero difference in propensity scores between the treatment and control group distributions further supports the effectiveness of our matching approach.

\subsection*{Difference-in-Differences Analysis Across Bike Lane Type}

We begin this section of the analysis with our dataset of treated units and their control matches. 
A station’s treatment indicator is always 0 or 1, but the post indicator is only set to 1 for treated units once the installation date (month, year) of the relevant adjacent bike lane is reached. The post indicator for control units remains 0 throughout. Our aim is to assess the average treatment effect, a measure of causality, that implementing a certain type of bike lane has on a nearby station’s ridership. In order to find the heterogeneous effects across `protected bike lane' and `painted bike lane', we turn to a two-way fixed effects difference-in-differences model with time- and unit-fixed effects:
\begin{equation}
Y_{it} = \alpha + \beta_0 W_{it} \cdot \ind(C_i = 0)
 + \beta_1 W_{it} \cdot \ind(C_i = 1)+ u_i + \omega_t + \epsilon_{it}
\end{equation}
where $Y_{it}$ represents the number of rides starting from station $i$ in month $t$. $W_{it}$ is a binary dummy variable that is 1 if station $i$ is treated and the installation year and month have been reached. 
The variable $C_i \in \{0,1\}$ denotes the type of bike lane treatment assigned to station $i$, with $C_i = 0$ indicating a painted bike lane and $C_i = 1$ indicating a protected bike lane. The coefficient $\beta_0$ thus captures the average treatment effect of painted bike lanes on ridership, while $\beta_1$ captures the effect of protected bike lanes. 
$u_i$ accounts for the station-level fixed effects, and $\omega_t$ represents time fixed effects (year and month). Finally, $\epsilon_{i,t}$ represents the error term. 
We see that the causal effect of bike lane implementation on ridership is reserved for protected lanes only (see Figure 2). Supplementary Table S2 shows the full regression results. 

\subsection*{Placebo Tests}
To assess the robustness of our difference-in-differences (DiD) estimates, we conducted a placebo test by randomly shuffling the installation dates of the bike lanes while keeping all other data fixed. For each of the 500 iterations, we reassigned treatment timing at random and re-estimated the DiD model using the same specification as in the main analysis. This process generated a distribution of placebo DiD estimates under the null hypothesis of no treatment effect. We then computed the 95th percentile of this null distribution as a benchmark for statistical significance. This placebo-based threshold allows us to evaluate whether the observed treatment effect is likely to have arisen by chance due to time-varying confounders or unobserved shocks. Figures S3 and S4 show the placebo tests for the protected and painted bike lanes, respectively. The results show that the effect is insignificant, supporting the claim that our results are not confounded by unobserved variables.

\subsection*{Difference-in-Differences Across Sociodemographic Variables}

Having demonstrated the causal effect of bike lane implementation on ridership as being isolated to protected lanes only, we aim to analyze disparities of this effect across sociodemographic variables. Examining only our 126 protected bike lane stations and their corresponding matches, we categorize each of these stations into low, medium, and high bins for each variable we are interested in, with the same number of units assigned to each category. Binning in this manner makes the model more sensitive to nonlinear terms, allowing more flexibility than a regression directly on the variable. The data for this analysis is the same data we used to perform PSM, with the addition of age data also taken from the 2023 ACS on the CBG level. We seek to assess heterogeneous effects within the following variables: median household income, percent [of the population] who identifies as black, percent 60-79 years of age, percent who identifies as Hispanic, and percent who has attained at least a bachelor’s degree. Correlations between these variables are shown in Supplementary Figure S5. 
We use the following DiD model:
\begin{equation}
Y_{it} = \alpha + \sum_{{b}} \beta_{b} W_{it} \cdot \ind(b_i = b) + u_i + \omega_t + \epsilon_{it}
\end{equation}
where $Y_{it}$ is the number of rides starting from station $i$ in month $t$, and $W_{it}$ is a binary indicator for whether a protected bike lane has been installed for station $i$ as of week $t$. The variable $b_i$ denotes the bin assignment for station $i$ based on the value of the moderating variable, categorized into three levels: low, medium, and high. The term $\mathbf{1}\{b_i = b\}$ is an indicator function equal to 1 if station $i$ falls into bin $b$, and 0 otherwise. The coefficient $\beta_b$ therefore represents the average treatment effect of protected bike lanes for stations in bin $b$.



Results in Figures 3 and 4, and Supplementary Tables S3 to S7 show that only the high bin for `percent Black' saw a causal effect on ridership. The reverse trend is found for the `percent aged 60-79' variable, with stations in CBGs with older populations being disproportionately causally impacted. 
The other three variables did not yield significant results when compared with the placebo experiments, shown in Figures S6 to S15. Because of the nature of the correlations between these five variables, more confidence can be assigned to the causal relationship between the two demographic factors which were found to be significant and station ridership.


\section*{Data Availability}
Both Citi Bike trip data and New York City’s bike-lane network are available for download from Citi Bike (https://Citi Bikenyc.com/system-data), and the NYC Open Data Portal (https://opendata.cityofnewyork.us/), respectively.

\bibliography{sample}



\section*{Author contributions statement}


M.M. and T.Y. conceptualized the work. M.S. and T.Y. designed the methodology, and designed and implemented the experiments. M.M. wrote the original draft, which was reviewed and edited by M.S. and T.Y. M.S. curated the data, which were visualized by M.M. and M.S. T.Y. supervised and administered the project.

\end{document}


\maketitle

\tableofcontents
\newpage

\listoffigures

\listoftables

\newpage

\setcounter{figure}{0}
\setcounter{table}{0}



\section{Data}
\subsection{Citi Bike Data}
Citi Bike data is available for download for all trips in the system's history, beginning in 2013. Trips include start date and time, end date and time, as well as start and end station. A number of other variables either were provided at one point and since discontinued (such as rider gender), or have been subsequently added and are not present in previous years (such as bike electrification status). Across years, station names have remained almost entirely consistent. Filtering Citi Bike trip data by unique station name allows the extent of all stations to be mapped for a given point in time. Citi Bike trip data can be downloaded from the following link: \url{https://citibikenyc.com/system-data}.  

\subsection{New York City Bike Lanes}
The City of New York publishes a map and dataset of its current bike lane network on its open-data portal, which can be downloaded and analyzed within GIS platforms. Each bike-lane segment includes installation month and year, as well as `facility class,' an attribute which reflects whether the segment is protected, painted, or a sharrow. All bike lane data for the analyses undertaken here reflect the city's network at the close of 2024. The bike lane network data of New York City can be downloaded from the following link: 
\url{https://data.cityofnewyork.us/dataset/New-York-City-Bike-Routes-Map-/9e2b-mctv}.

\subsection{Correlational Analysis of Citi Bike Trips and Bike-Lane Types}
After loading both unique Citi Bike Stations and New York's bike lane segments into a GIS platform, a spatial join was conducted such that any bike lane segments within 100 meters (as the crow flies) of a Citi Bike station were joined. Taking into account the installation month and year of each bike-lane segment, as well as bike-lane type (combining painted bike lanes and sharrows), that allowed for a comparison of Citi Bike trips for the 12 months prior to installation, to the 12 months after, which were averaged across types.  

\section{Causal Inference Framework}

\subsection{Treatment Assignment}
We begin with a dataset of all Citi Bike stations and their corresponding bike lanes (within 100 meters). Since Citi Bike trip data is available from 2013 to 2024, we make sure that those units classified as treated had nearby bike instillation/s after 2013 and before 2024. This leaves at least one year before and one year after the installation for our difference-in-differences (DiD) analysis. If the station had at least one protected lane installed nearby, it is classified as a treated unit with a protected bike lane. If the station did not have a protected lane installed, but instead saw a painted intervention (this includes both painted lines and sharrows) implemented, the bike station is labeled as a treated unit but with a painted bike lane. If only one installation took place, this installation’s date is used for analysis. If more than one occurred, the earliest installation date of the relevant class is used. 

A similar approach is taken to identify eligible control units. The filter now selects installation dates that fall before 2014, as well as stations that never had bike lanes installed nearby (untreated across all time). As before, stations are classified into categories: treated with protected bike lane, treated with painted bike lane, and untreated. Untreated in this context means not treated in or between 2014 and 2023. In the event that a station had a painted lane installed during the treatment period, but was already in the protected lane type before this period, it was considered untreated. And vice versa: a painted lane station upgraded to a protected lane station during the treatment period is considered treated. 

Once treated and control groups were selected, we merged the classified stations with the trip data. We retrieve the first month and year that a station saw ridership and keep only those treated stations where the station came before the bike lane installation. This ensures that there is ridership to compare pre- and post-install. Lastly, we take into account extended station closures (due to construction, maintenance, etc.) by dropping stations that have more than three consecutive zero-ridership months after ridership first appears. We end this section of the methods with 240 treated units and 1,406 eligible control stations.

\subsection{Propensity Score Matching}
Before performing the formal difference-in-differences (DiD) analysis, we refine the selection of control stations using propensity score matching (PSM) with replacement. This approach allows each control station to be matched with multiple treated POIs, increasing the robustness of our analysis. We focus on one-to-one matching with replacement as it generally leads to greater bias reduction than one-to-one matching without replacement. Specifically, we employ nearest-neighbor PSM to match treated POIs with control POIs that exhibit similar characteristics. The matching criteria are the following: median household income, percent of the population that identifies as black and hispanic, percent that has attained at least a bachelor’s degree, stations within 200m, dining establishments within 200m, and population density.

In order to perform PSM, we utilize several data sources. First, we leverage 2023 American Community Survey (ACS) data for the state of New York — this provides us with census-block-group (CBG) level data on median household income, total population, black population, hispanic population, and educational attainment. The last of these references the population of people with degrees equal to, or more advanced than, a bachelor's. We use the CBG information from the U.S. Census Bureau's TIGER/Line Shapefiles interface. Using this information, we are able to calculate population density, as well as sociodemographic percentages relative to total population.

Another covariate we account for is the density of POIs. From Foursquare’s POI data, we isolate dining establishments specifically, and then perform a k-d tree query to locate the number of dining establishments within 200 meters of each station. We use a similar approach to calculate network density: using a k-d tree model, we identify the number of Citi Bike stations within 200 meters of each station. Lastly, we exclude those CBGs where population is equal to zero and impute median household income for those CBGs for which it is absent from the dataset (this occurs 252 times, but never for a treated unit). We end this part of the process with 222 treated units and 1,318 eligible control units.

To examine the relationship between our covariates and the likelihood of treatment, we fit a logistic regression model where the treatment variable is regressed on the covariates:

\begin{align}
\log\left(\frac{P(\text{Treated} = 1)}{1 - P(\text{Treated} = 1)}\right) 
&= \beta_0 
+ \sum_{i=1}^{g} \beta_{1,i} \cdot \text{Population Density}_i \nonumber \\
&\quad + \beta_2 \cdot \text{Median Household Income} 
+ \beta_3 \cdot \text{Percent Black} \nonumber \\
&\quad + \beta_4 \cdot \text{Percent Hispanic} 
+ \beta_5 \cdot \text{Dining Within 200m} \nonumber \\
&\quad + \beta_6 \cdot \text{Stations Within 200m}
\end{align}

This model has an accuracy score of 0.885, demonstrating reasonable predictive performance. Detailed regression results are presented in Supplementary Table S1. Once a propensity score is assigned to each Citi Bike station, we turn to nearest neighbor matching to match each treated unit with a control unit.

\begin{table}[htbp]
\centering
\caption{Logistic Regression Predicting Treatment Assignment}
\begin{tabular}{l@{\hskip 2em}c}
\toprule
 & Coefficient \\
\midrule
Intercept & $-1.81^{***}$ \\
          & (0.24), $p < 0.001$ \\
Median Household Income & $1.31 \times 10^{-7}$ \\
          & (2.00e-6), $p = 0.948$ \\
Percent Black & $-0.61$ \\
          & (0.38), $p = 0.104$ \\
Bachelor’s Degree (\%) & $0.0033$ \\
          & (0.005), $p = 0.471$ \\
Population Density & $-1.26$ \\
          & (3.88), $p = 0.744$ \\
Stations within 200m & $-0.73^{***}$ \\
          & (0.17), $p < 0.001$ \\
Dining Options within 200m & $0.0035^{***}$ \\
          & (0.001), $p < 0.001$ \\
\addlinespace
\midrule
Observations & 1,542 \\
Pseudo $R^2$ & 0.037 \\
Log-Likelihood & $-601.66$ \\
LL-Null & $-624.67$ \\
Likelihood Ratio $p$-value & $< 0.001$ \\
\bottomrule
\end{tabular}
\begin{flushleft}
\footnotesize
\textit{Notes:} Standard errors in parentheses. $^{***}p<0.01$, $^{**}p<0.05$, $^{*}p<0.1$
\end{flushleft}
\end{table}

After the matching process, we assess the covariate balance between the treatment and control groups to ensure comparability. The standardized mean differences (SMD) plot in Supplementary Figure S1 and the propensity score distribution shown in Supplementary Figure S2, confirm the achievement of a balanced covariate distribution between the groups. More specifically, The SMD measures the difference in means in the unit of pooled standard deviation for a specific covariate, and is calculated as follows:

\begin{equation}
\text{SMD} = \frac{\bar{x}_{treatment} - \bar{x}_{control}}{\sqrt{\frac{s_{treatment}^2 + s_{control}^2}{2}}}
\end{equation}
where $\bar{x}_{treatment}$ and $\bar{x}_{control}$ denote the sample mean of the covariate in treated and untreated subjects, respectively, whereas $s_{treatment}^2$ and $s_{control}^2$ denote the sample variance of the covariate in treated and untreated subjects, respectively. An SMD below 0.10 for a given covariate suggests sufficient overlap between the treatment and control groups, indicating that the matching process has successfully minimized bias in the covariate distributions. Furthermore, the near-zero difference in propensity scores between the treatment and control group distributions further supports the effectiveness of our matching approach.

\begin{figure}[h]
    \centering
    \begin{subfigure}[b]{0.7\linewidth}
        \centering
        \includegraphics[width=\linewidth]{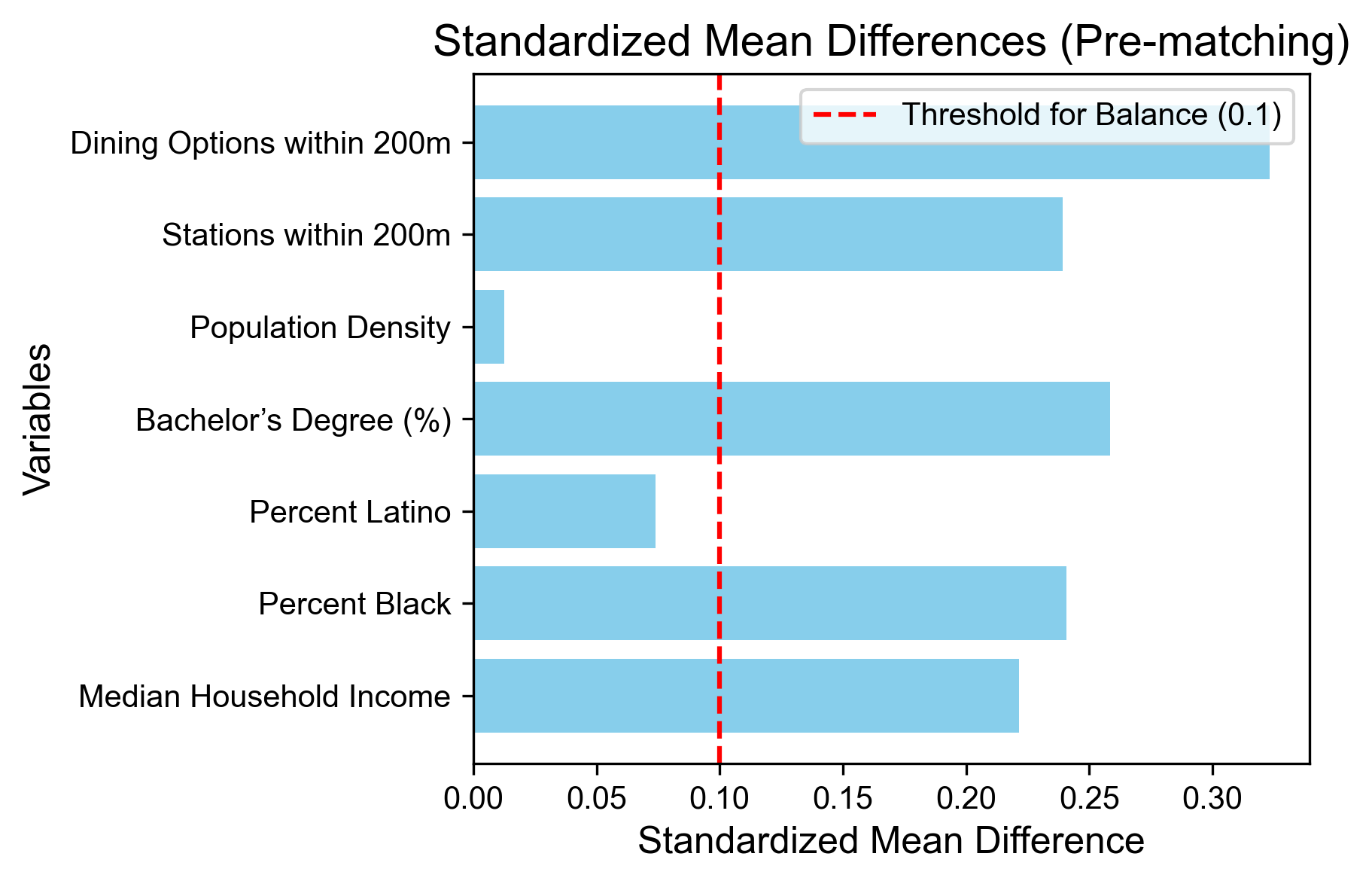}
        \caption{\textbf{Standard mean difference before matching}}
        \label{fig:smd_before}
    \end{subfigure}
    
    \vspace{1em}  
    
    \begin{subfigure}[b]{0.7\linewidth}
        \centering
        \includegraphics[width=\linewidth]{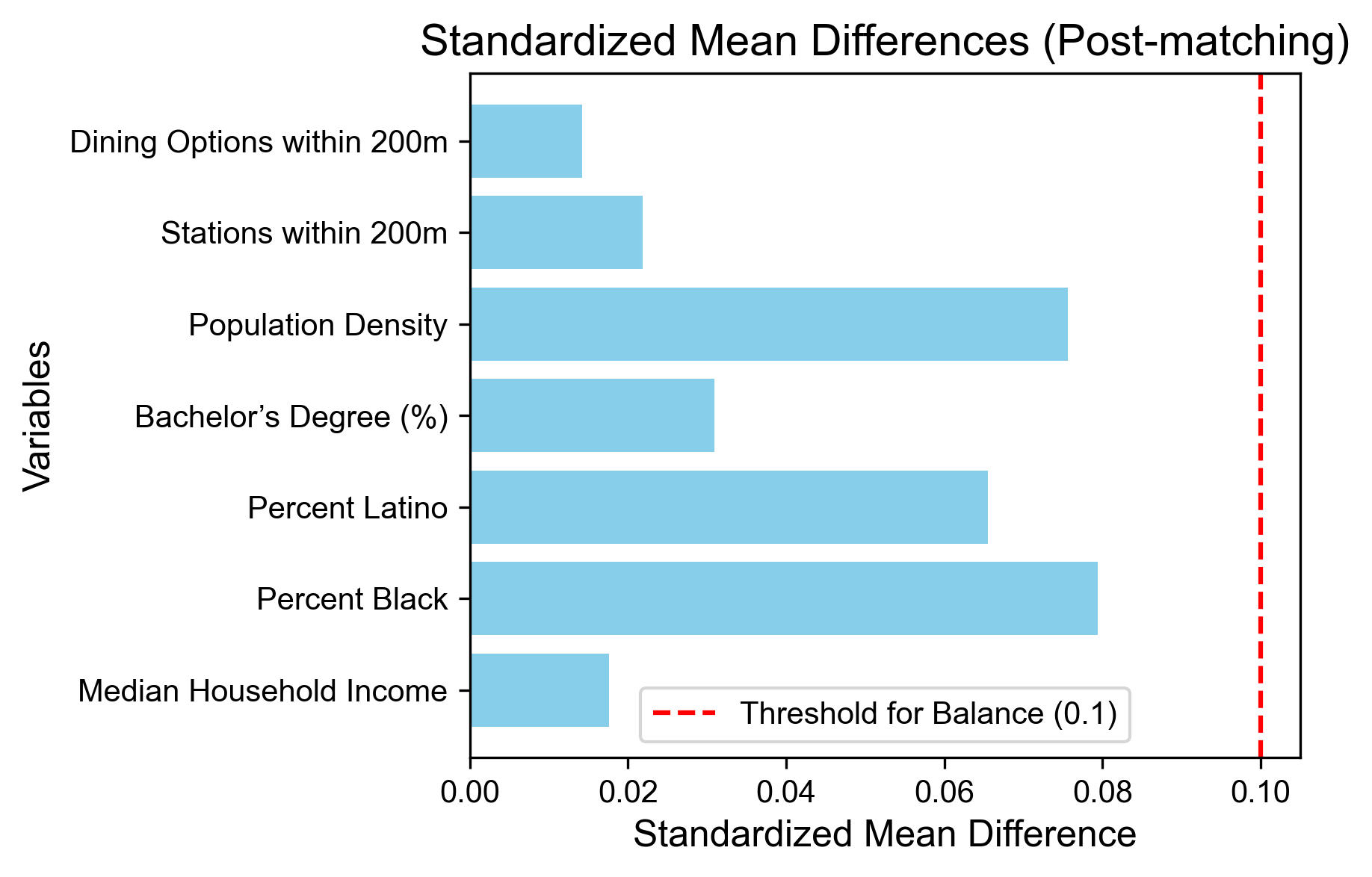}
        \caption{\textbf{Standard mean difference after matching}}
        \label{fig:smd_after}
    \end{subfigure}
    
    \caption{Standardized mean differences before and after matching.}
    \label{fig:smd_combined}
\end{figure}

\begin{figure}[h]
    \centering
    \begin{subfigure}[b]{0.8\linewidth}
        \centering
        \includegraphics[width=\linewidth]{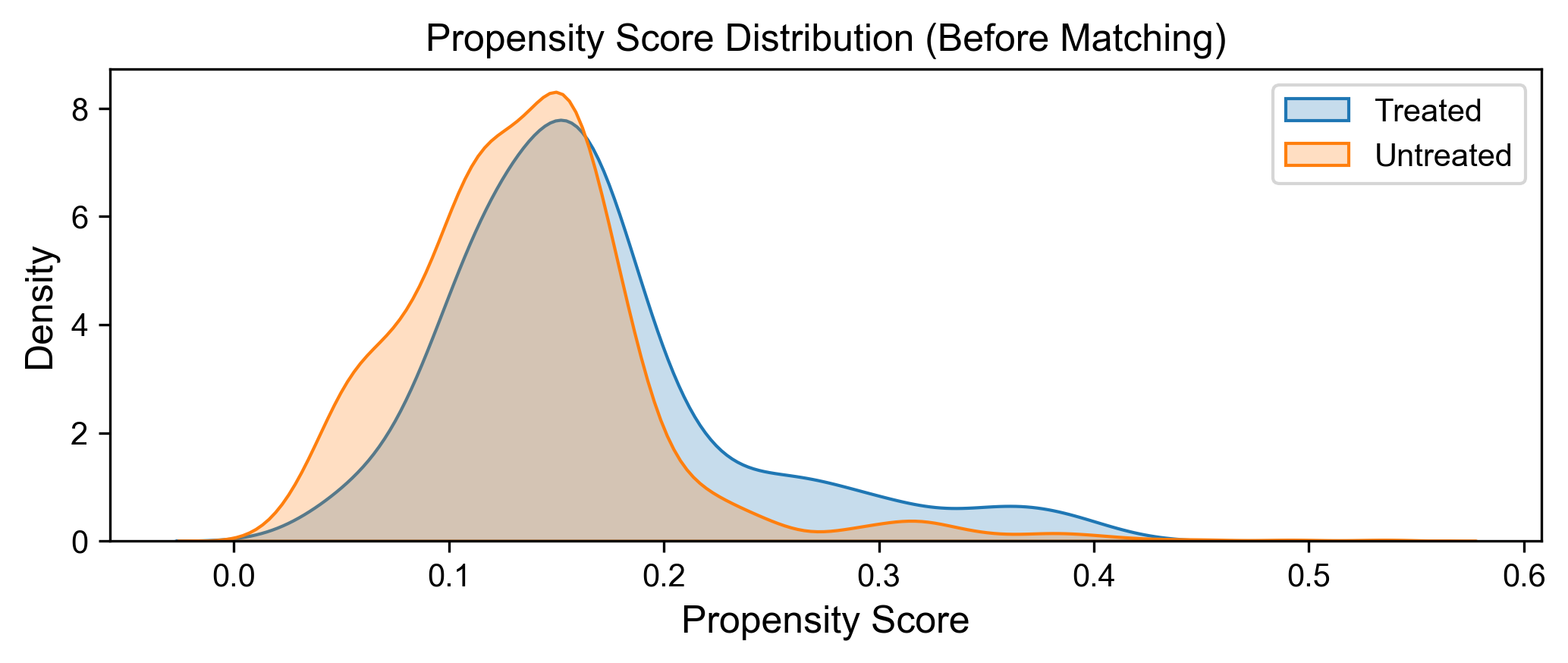}
        \caption{\textbf{Propensity scores before matching}}
        \label{fig:smd_before}
    \end{subfigure}
    
    \vspace{1em}  
    
    \begin{subfigure}[b]{0.8\linewidth}
        \centering
        \includegraphics[width=\linewidth]{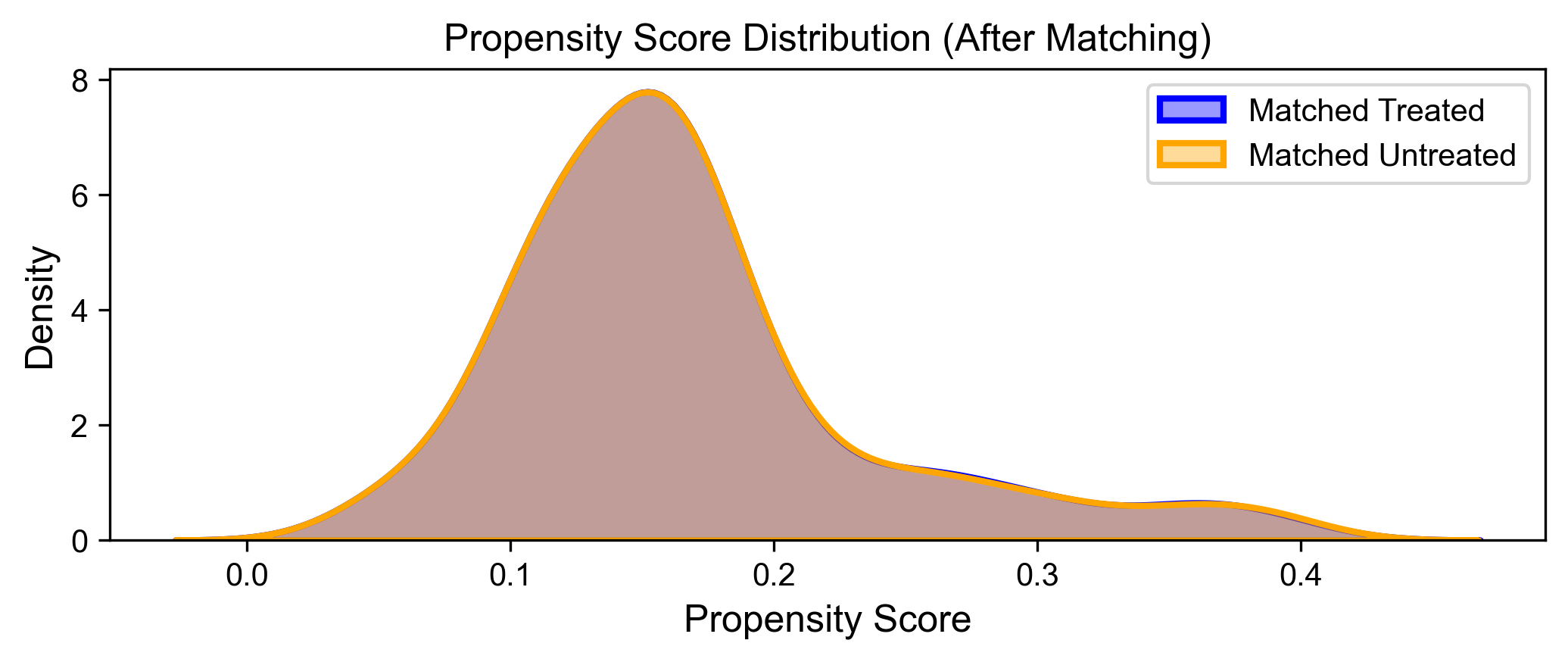}
        \caption{\textbf{Propensity scores after matching}}
        \label{fig:smd_after}
    \end{subfigure}
    
    \caption{Propensity scores before and after matching.}
    \label{fig:smd_combined}
\end{figure}

\subsection{Difference in Differences Analysis Across Bike Lane Type}

We begin this section of the analysis with our dataset of treated units and their control matches. 
A station’s treatment indicator is always 0 or 1, but the post indicator is only set to 1 for treated units once the installation date (month, year) of the relevant nearby bike lane is reached. The post indicator for control units remains 0 throughout. Our aim is to assess the average treatment effect, a measure of causality, that implementing a certain type of bike lane has on a nearby station’s ridership. In order to find the heterogeneous effects across `protected bike lane' and `painted bike lane', we turn to a two-way fixed effects difference-in-differences model with time- and unit-fixed effects:

\begin{equation}
Y_{it} = \alpha + \beta_0 W_{it} \cdot \ind(C_i = 0)
 + \beta_1 W_{it} \cdot \ind(C_i = 1)+ u_i + \omega_t + \epsilon_{it}
\end{equation}
where $Y_{it}$ represents the number of rides starting from station $i$ in month $t$. $W_{it}$ is a binary dummy variable that is 1 if station $i$ is treated and the installation year and month have been reached. 
The variable $C_i \in \{0,1\}$ denotes the type of bike lane treatment assigned to station $i$, with $C_i = 0$ indicating a painted bike lane and $C_i = 1$ indicating a protected bike lane. The coefficient $\beta_0$ thus captures the average treatment effect of painted bike lanes on ridership, while $\beta_1$ captures the effect of protected bike lanes. 
$u_i$ accounts for the station-level fixed effects, and $\omega_t$ represents time fixed effects (year and month). Finally, $\epsilon_{i,t}$ represents the error term. 
By allowing separate coefficients for the two categories, the model accommodates heterogeneous treatment effects by infrastructure type.

We run this two way fixed effects DiD model to obtain the causal estimates $\beta_0$ and $\beta_1$, which are the average treatment effects for each treatment class.
We see that the causal effect of bike lane implementation on ridership is reserved for protected lanes only (see Figure 2). Supplementary Table S2 shows the full regression results. 

\begin{table}[htbp]
\centering
\caption{Effect of Treatment on Bike Trips for both protected and painted bike lanes}
\begin{tabular}{l@{\hskip 2em}c}
\toprule
 & (1) \\
\midrule
Constant & $3432.43^{***}$ \\
 & (67.17), $p < 0.001$ \\
Treated × Post × Protected Bike Lane & $379.88^{***}$ \\
 & (123.46), $p < 0.001$ \\
Treated × Post × Painted Bike Lane & $-141.18$ \\
 & (99.64.98), $p = 0.156$ \\
\addlinespace
Bike Station Fixed Effects & Yes \\
Year-Month Fixed Effects & Yes \\
\midrule
Observations & 57,024 \\
$R^2$ & 0.753 \\
Adjusted $R^2$ & 0.750 \\
\bottomrule
\end{tabular}
\begin{flushleft}
\footnotesize
\textit{Notes:} Robust standard errors clustered at the station level in parentheses. \\
$^{***}p<0.01$, $^{**}p<0.05$, $^{*}p<0.1$
\end{flushleft}
\end{table}


\subsection{Placebo Tests}
To assess the robustness of our difference-in-differences (DiD) estimates, we conducted a placebo test by randomly shuffling the installation dates of the bike lanes while keeping all other data fixed. For each of the 500 iterations, we reassigned treatment timing at random and re-estimated the DiD model using the same specification as in the main analysis. This process generated a distribution of placebo DiD estimates under the null hypothesis of no treatment effect. We then computed the 95th percentile of this null distribution as a benchmark for statistical significance. This placebo-based threshold allows us to evaluate whether the observed treatment effect is likely to have arisen by chance due to time-varying confounders or unobserved shocks. Figures S3 and S4 show the placebo tests for the protected and painted bike lanes, respectively. The results show that the effect is insignificant, supporting the claim that our results are not confounded by unobserved variables. 

\begin{figure}[h]
    \centering
    \includegraphics[width=0.8\linewidth]{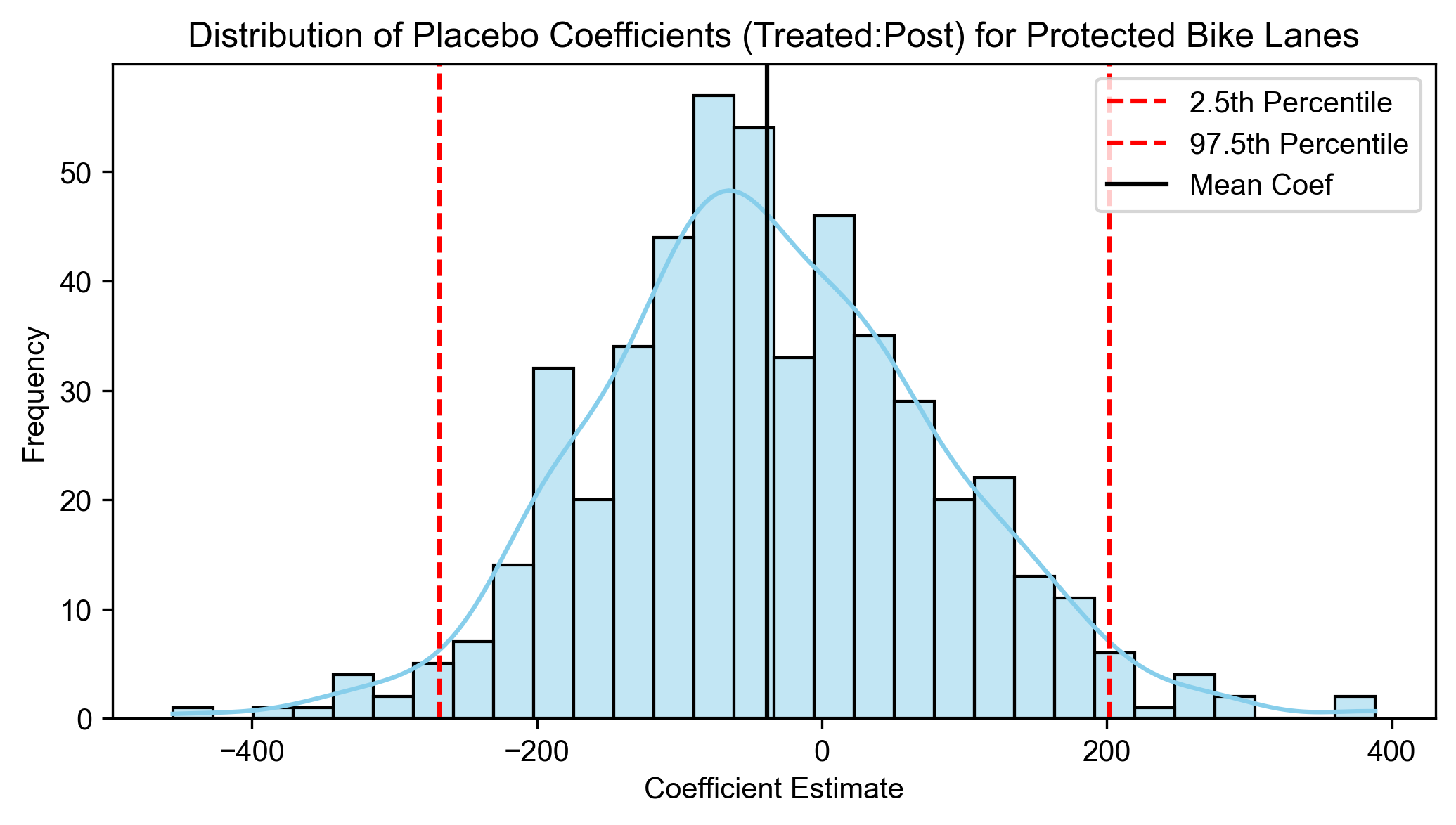}
    \caption[DiD coefficients for placebo test (Protected bike lanes)]{\textbf{DiD coefficients for placebo test (Protected bike lanes).}}
    \label{fig:Figure S1}
\end{figure}

\begin{figure}[h]
    \centering
    \includegraphics[width=0.8\linewidth]{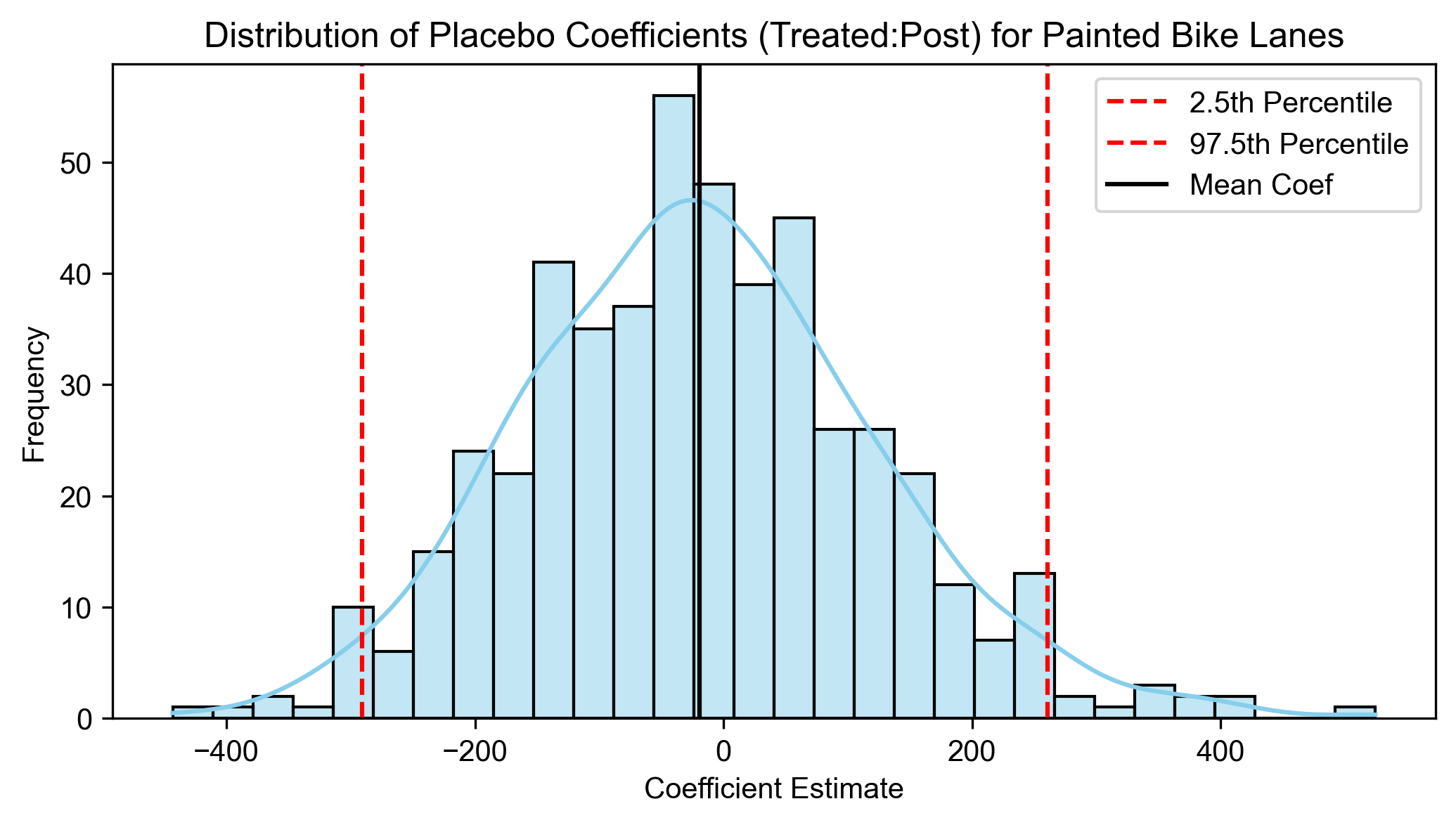}
    \caption[DiD coefficients for placebo test (Painted bike lanes)]{\textbf{DiD coefficients for placebo test (Painted bike lanes).}}
    \label{fig:Figure S1}
\end{figure}

\section{Heterogeneous Difference in Differences Analysis Across Sociodemographic Variables}
Having demonstrated the causal effect of bike lane implementation on ridership as being isolated to protected lanes only, we aim to analyze disparities of this effect across sociodemographic variables. Examining only our 132 protected bike lane stations and their corresponding matches, we categorize each of these stations into low, medium, and high bins for each variable we are interested in, with the same number of units assigned to each category. The data for this analysis are the same data we used to perform PSM, with the addition of age data also taken from the 2023 American Community Survey (ACS) on the CBG level. We seek to assess heterogeneous effects within the following variables: median household income, percent [of the population] that identifies as black, percent older than 60 years of age, percent that identifies as Hispanic, and percent that has attained at least a bachelor’s degree. Correlations between these variables are shown in Supplementary Figure S5. 
We use the following DiD model:

\begin{equation}
Y_{it} = \alpha + \sum_{{b}} \beta_{b} W_{it} \cdot \ind(b_i = b) + u_i + \omega_t + \epsilon_{it}
\end{equation}
where $Y_{it}$ is the number of rides starting from station $i$ in month $t$, and $W_{it}$ is a binary indicator for whether a protected bike lane has been installed for station $i$ as of month $t$. The variable $b_i$ denotes the bin assignment for station $i$ based on the value of the moderating variable, categorized into three levels: low, medium, and high. The term $\mathbf{1}\{b_i = b\}$ is an indicator function equal to 1 if station $i$ falls into bin $b$, and 0 otherwise. The coefficient $\beta_b$ therefore represents the average treatment effect of protected bike lanes for stations in bin $b$.

\begin{figure}[h]
    \centering
    \includegraphics[width=0.75\linewidth]{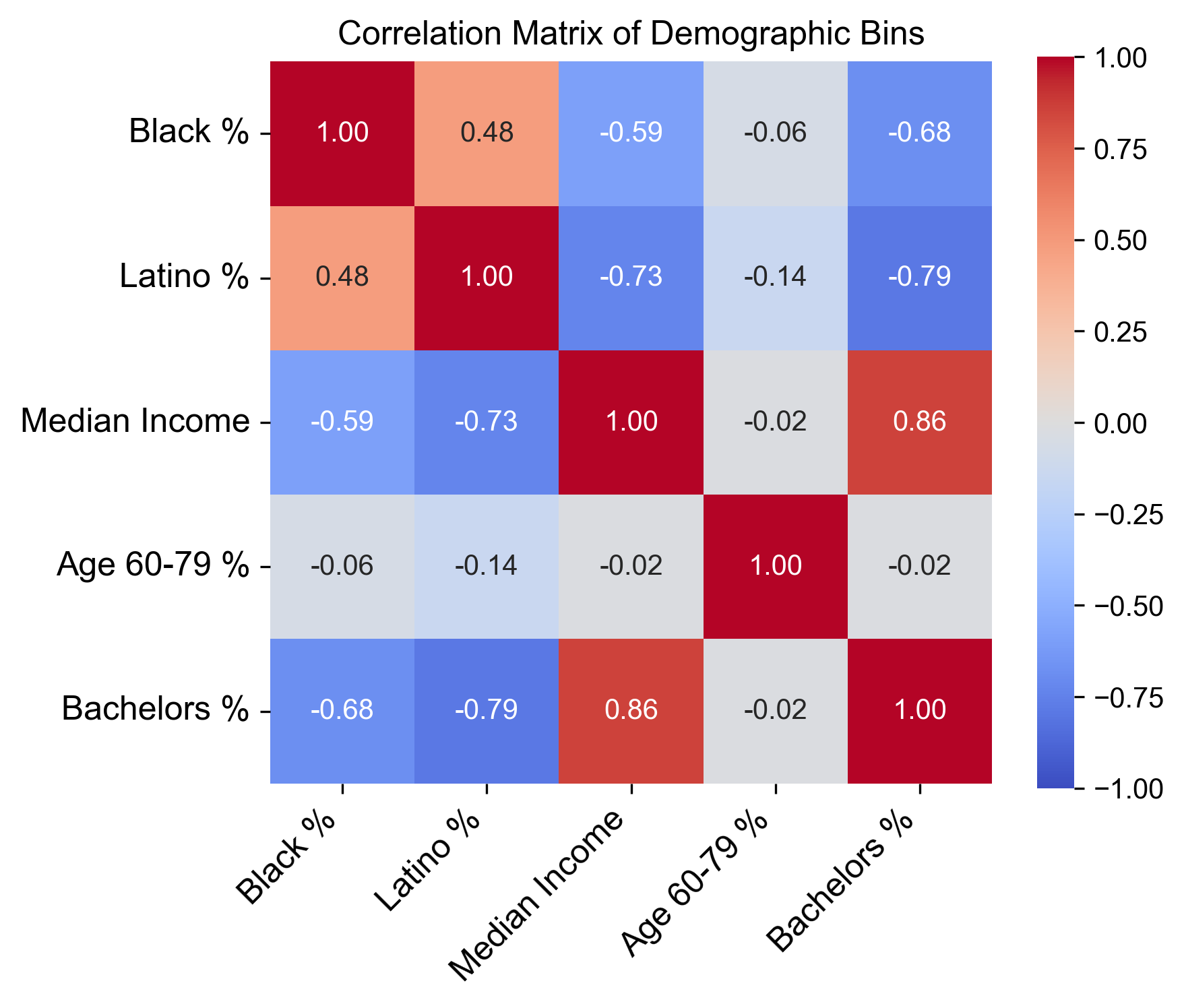}
    \caption[Correlation matrix of sociodemographic variables]{\textbf{Correlation matrix of sociodemographic variables.} Pearson correlation between sociodemographic variables in census block groups where Citi Bike stations are located. Black \%, Hispanic \%, median household income, and \% of Bachelor degrees are all correlated with eachother. \% of age 60 -- 79 is not stronly correlated with any of the variables.}
    \label{fig:Figure S1}
\end{figure}



Results in Figures 3 and 4, and Supplementary Tables S3 to S7 show that only the high bin for `percent black' saw a causal effect on ridership. The reverse trend is found for the `percent aged 60+' variable, with stations in CBGs with older populations being disproportionately causally impacted. 
The other three variables did not yield significant results when compared with the placebo experiments, shown in Figures S6 to S15. Because of the nature of the correlations between these five variables, more confidence can be assigned to the causal relationship between the two demographic factors which were found to be significant.




\begin{table}[htbp]
\centering
\caption{Effect of Treatment on Bike Trips by Percent Black}
\begin{tabular}{l@{\hskip 2em}c}
\toprule
 & (1) \\
\midrule
Constant & $3267.02^{***}$ \\
 & (95.24), $p < 0.001$ \\
Percent Black  (Low) × Treated × Post & $464.00^{**}$ \\
 & (181.59), $p = 0.011$ \\
Percent Black (Medium) × Treated × Post & $271.80$ \\
 & (186.21), $p = 0.144$ \\
Percent Black (High) × Treated × Post & $80.68$ \\
 & (248.11), $p = 0.745$ \\
\addlinespace
Bike Station Fixed Effects & Yes \\
Year-Month Fixed Effects & Yes \\
\midrule
Observations & 33,792 \\
$R^2$ & 0.758 \\
Adjusted $R^2$ & 0.755 \\
\bottomrule
\end{tabular}
\begin{flushleft}
\footnotesize
\textit{Notes:} Robust standard errors clustered at the station level in parentheses. \\
$^{***}p<0.01$, $^{**}p<0.05$, $^{*}p<0.1$
\end{flushleft}
\end{table}

\begin{table}[htbp]
\centering
\caption{Effect of Treatment on Bike Trips by Percent Latino}
\begin{tabular}{l@{\hskip 2em}c}
\toprule
 & (1) \\
\midrule
Constant & $3270.66^{***}$ \\
 & (95.27), $p < 0.001$ \\
Percent Latino (Low) × Treated × Post & $684.19^{***}$ \\
 & (199.13), $p < 0.001$ \\
Percent Latino (Medium) × Treated × Post & $378.43^{*}$ \\
 & (227.72), $p = 0.097$ \\
Percent Latino (High) × Treated × Post & $-298.25^{**}$ \\
 & (149.13), $p = 0.046$ \\
\addlinespace
Bike Station Fixed Effects & Yes \\
Year-Month Fixed Effects & Yes \\
\midrule
Observations & 33,792 \\
$R^2$ & 0.760 \\
Adjusted $R^2$ & 0.758 \\
\bottomrule
\end{tabular}
\begin{flushleft}
\footnotesize
\textit{Notes:} Robust standard errors clustered at the station level in parentheses. \\
$^{***}p<0.01$, $^{**}p<0.05$, $^{*}p<0.1$
\end{flushleft}
\end{table}

\begin{table}[htbp]
\centering
\caption{Effect of Treatment on Bike Trips by Median Income}
\begin{tabular}{l@{\hskip 2em}c}
\toprule
 & (1) \\
\midrule
Constant & $3269.15^{***}$ \\
 & (95.41), $p < 0.001$ \\
Median Income (Low) × Treated × Post & $-205.77$ \\
 & (190.45), $p = 0.280$ \\
Median Income (Medium) × Treated × Post & $467.23^{**}$ \\
 & (210.22), $p = 0.026$ \\
Median Income (High) × Treated × Post & $533.61^{**}$ \\
 & (209.21), $p = 0.011$ \\
\addlinespace
Bike Station Fixed Effects & Yes \\
Year-Month Fixed Effects & Yes \\
\midrule
Observations & 33,792 \\
$R^2$ & 0.759 \\
Adjusted $R^2$ & 0.757 \\
\bottomrule
\end{tabular}
\begin{flushleft}
\footnotesize
\textit{Notes:} Robust standard errors clustered at the station level in parentheses. \\
$^{***}p<0.01$, $^{**}p<0.05$, $^{*}p<0.1$
\end{flushleft}
\end{table}

\begin{table}[htbp]
\centering
\caption{Effect of Treatment on Bike Trips by Age 60-79 \%}
\begin{tabular}{l@{\hskip 2em}c}
\toprule
 & (1) \\
\midrule
Constant & $3270.19^{***}$ \\
 & (95.60), $p < 0.001$ \\
Age 60-79 \% (Low) × Treated × Post & $-122.98$ \\
 & (182.80), $p = 0.501$ \\
Age 60-79 \% (Medium) × Treated × Post & $269.18$ \\
 & (187.94), $p = 0.152$ \\
Age 60-79 \% (High) × Treated × Post & $688.82^{***}$ \\
 & (231.62), $p < 0.001$ \\
\addlinespace
Bike Station Fixed Effects & Yes \\
Year-Month Fixed Effects & Yes \\
\midrule
Observations & 33,792 \\
$R^2$ & 0.759 \\
Adjusted $R^2$ & 0.757 \\
\bottomrule
\end{tabular}
\begin{flushleft}
\footnotesize
\textit{Notes:} Robust standard errors clustered at the station level in parentheses. \\
$^{***}p<0.01$, $^{**}p<0.05$, $^{*}p<0.1$
\end{flushleft}
\end{table}

\begin{table}[htbp]
\centering
\caption{Effect of Treatment on Bike Trips by Bachelors \%}
\begin{tabular}{l@{\hskip 2em}c}
\toprule
 & (1) \\
\midrule
Constant & $3269.55^{***}$ \\
 & (95.09), $p < 0.001$ \\
Bachelors \% (Low) × Treated × Post & $-131.24$ \\
 & (235.55), $p = 0.577$ \\
Bachelors \% (Medium) × Treated × Post & $104.02$ \\
 & (157.03), $p = 0.508$ \\
Bachelors \% (High) × Treated × Post & $828.93^{***}$ \\
 & (211.76), $p < 0.001$ \\
\addlinespace
Bike Station Fixed Effects & Yes \\
Year-Month Fixed Effects & Yes \\
\midrule
Observations & 33,792 \\
$R^2$ & 0.760 \\
Adjusted $R^2$ & 0.758 \\
\bottomrule
\end{tabular}
\begin{flushleft}
\footnotesize
\textit{Notes:} Robust standard errors clustered at the station level in parentheses. \\
$^{***}p<0.01$, $^{**}p<0.05$, $^{*}p<0.1$
\end{flushleft}
\end{table}

\begin{figure}[h]
    \centering
    \includegraphics[width=1\linewidth]{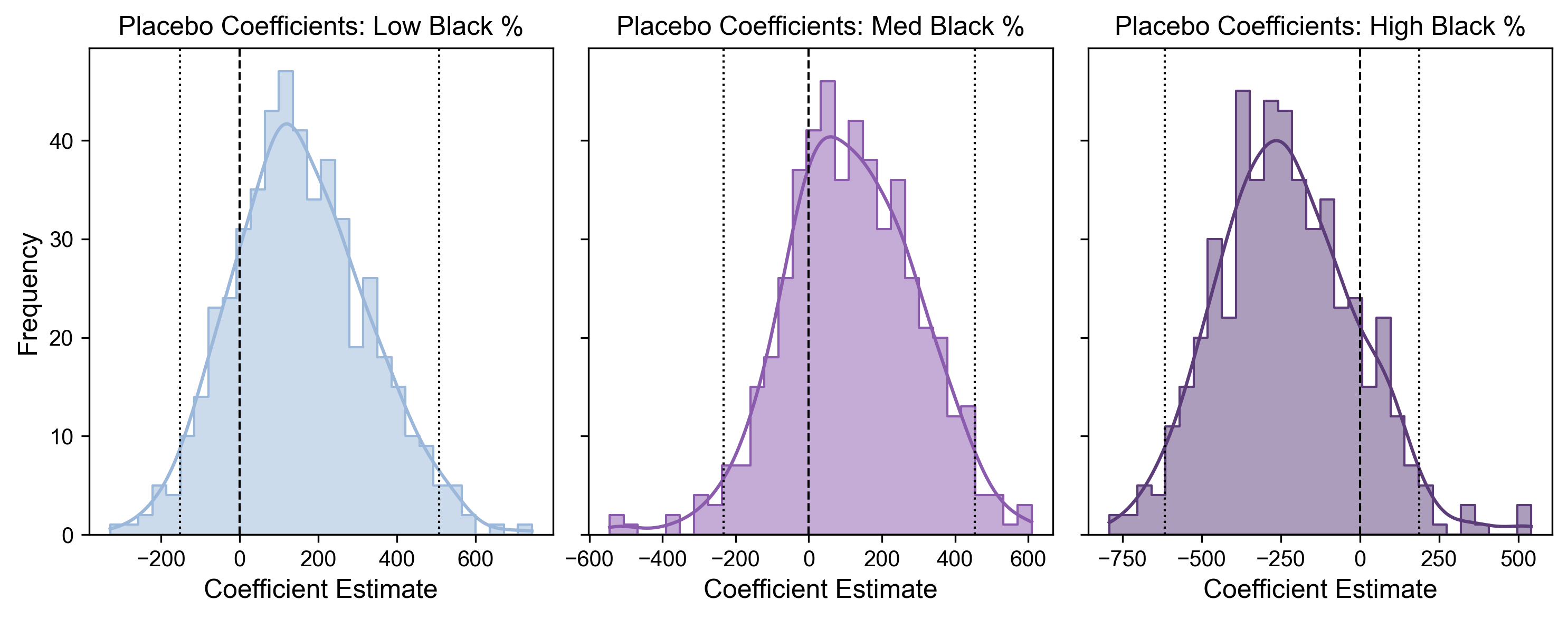}
    \caption[Placebo test coefficients for Black \%]{\textbf{Placebo test coefficients for Black \%.}}
    \label{fig:Figure S1}
\end{figure}

\begin{figure}[h]
    \centering
    \includegraphics[width=.65\linewidth]{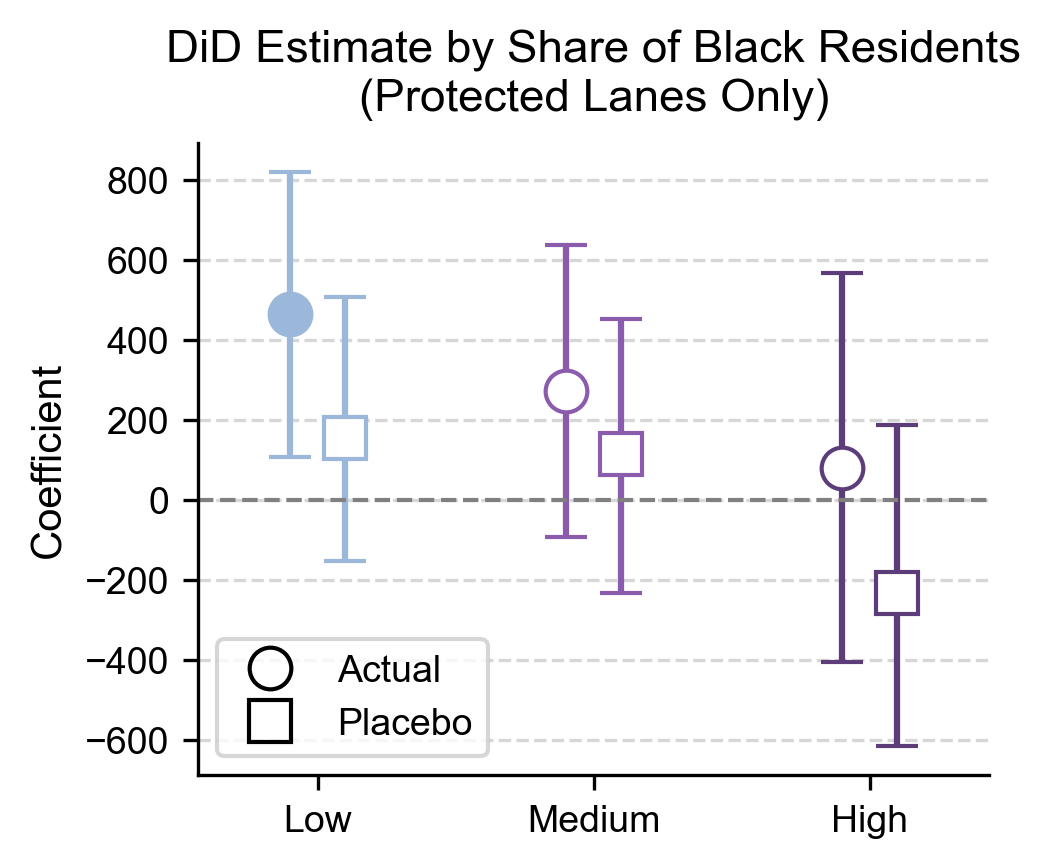}
    \caption[DiD results for Black \%]{\textbf{DiD results for Black \%.}}
    \label{fig:Figure S1}
\end{figure}

\begin{figure}[h]
    \centering
    \includegraphics[width=1\linewidth]{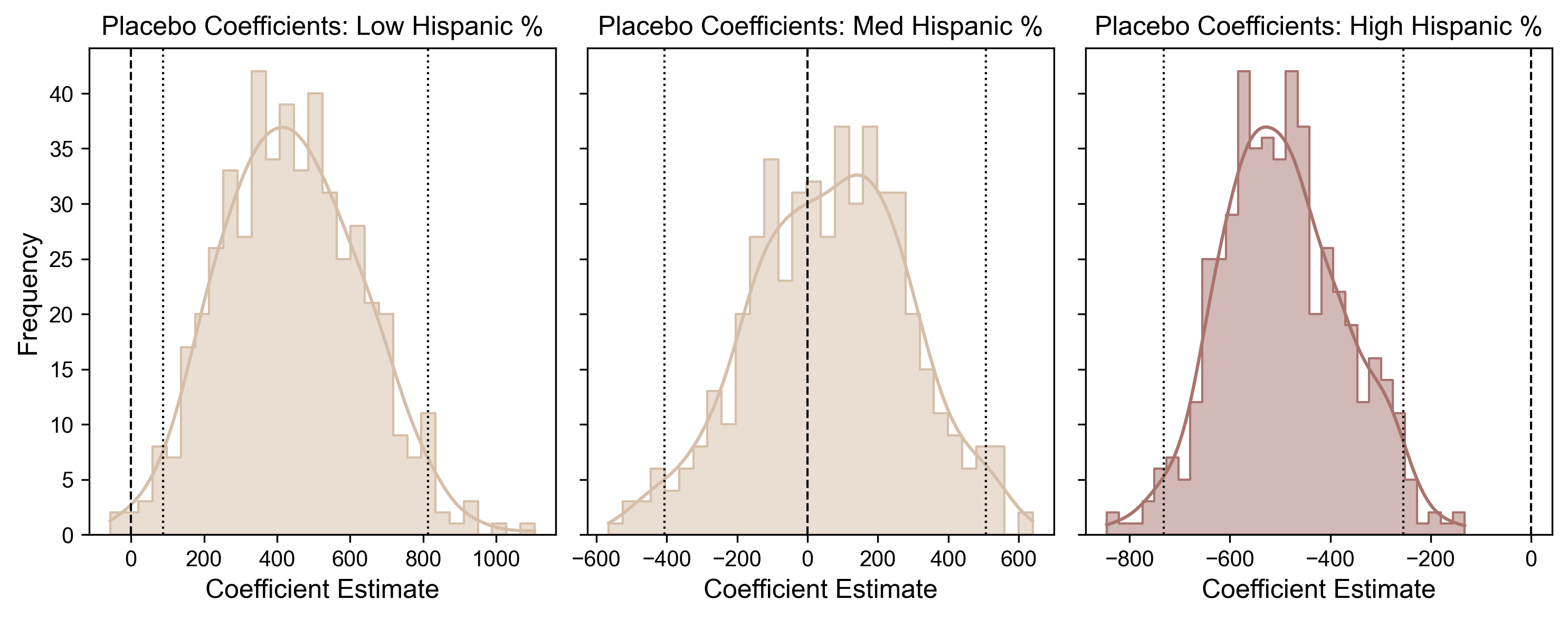}
    \caption[Placebo test coefficients for Hispanic \%]{\textbf{Placebo test coefficients for Hispanic \%.}}
    \label{fig:Figure S1}
\end{figure}

\begin{figure}[h]
    \centering
    \includegraphics[width=.65\linewidth]{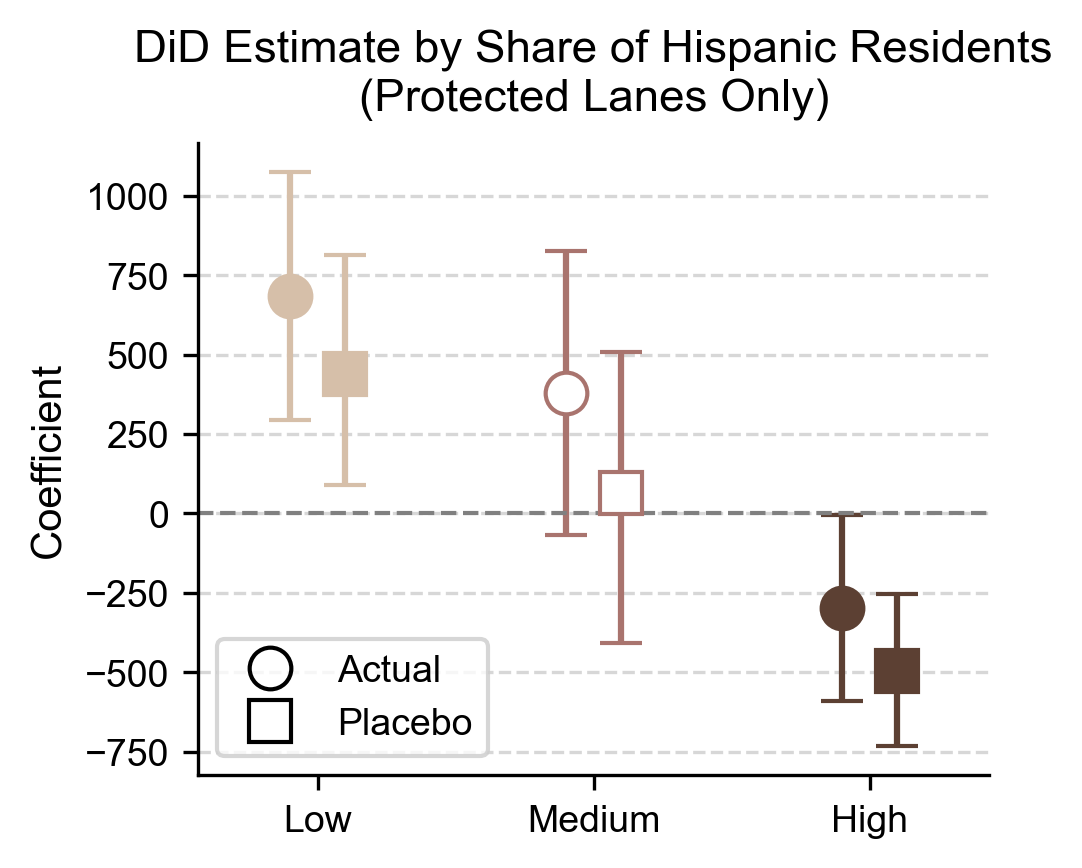}
    \caption[DiD results for Hispanic \%]{\textbf{DiD results for Hispanic \%.}}
    \label{fig:Figure S1}
\end{figure}

\begin{figure}[h]
    \centering
    \includegraphics[width=1\linewidth]{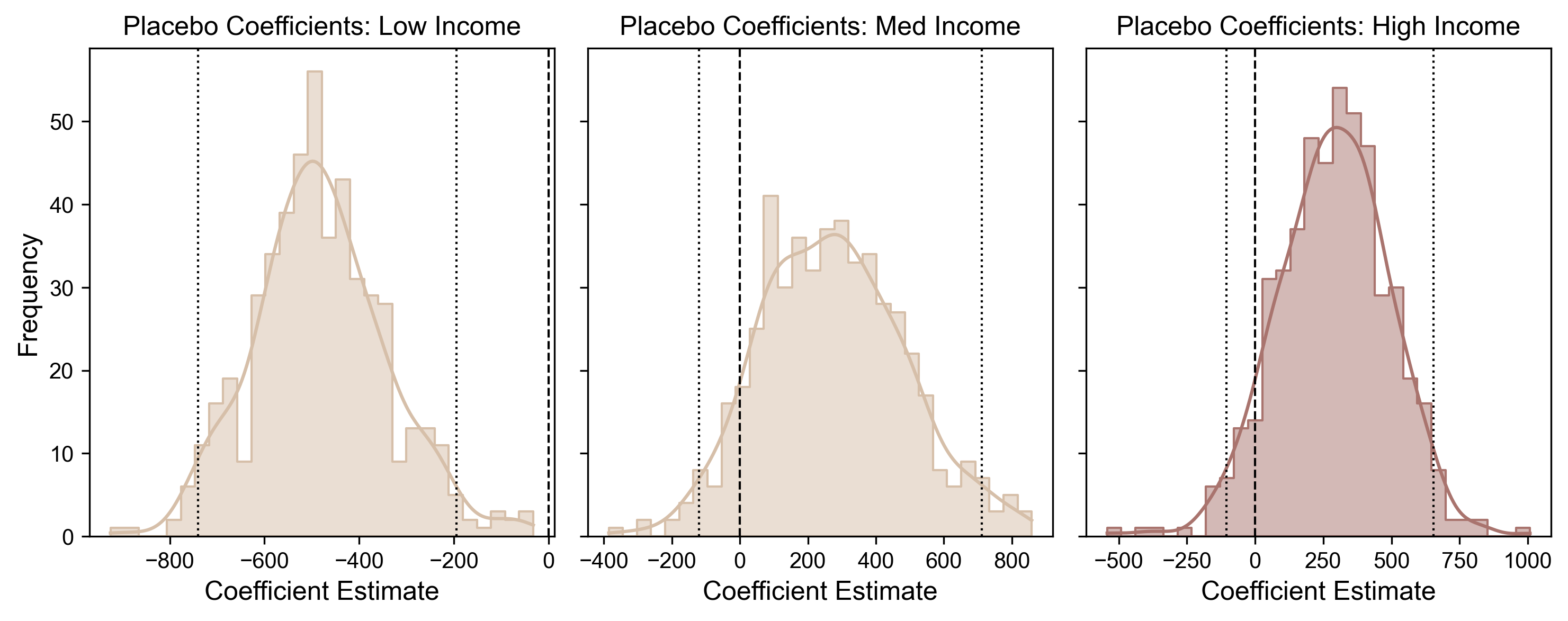}
    \caption[Placebo test coefficients for median income]{\textbf{Placebo test coefficients for median income.}}
    \label{fig:Figure S1}
\end{figure}

\begin{figure}[h]
    \centering
    \includegraphics[width=.65\linewidth]{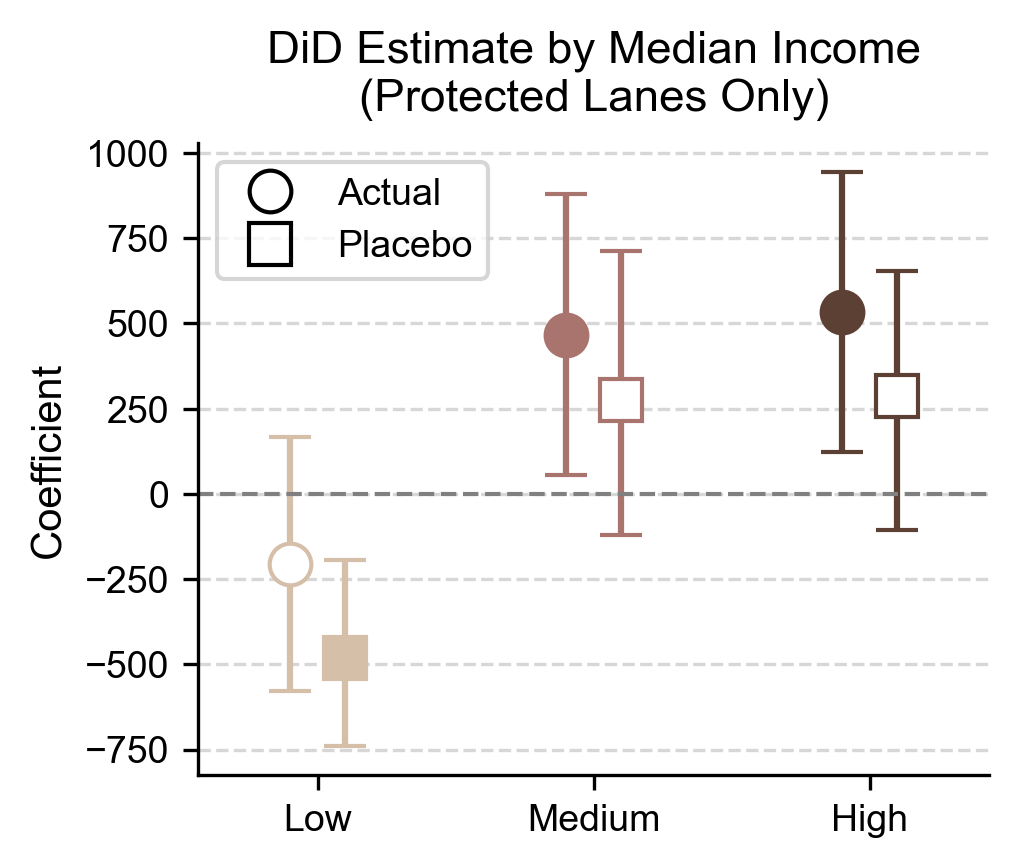}
    \caption[DiD results for median income]{\textbf{DiD results for median income.}}
    \label{fig:Figure S1}
\end{figure}

\begin{figure}[h]
    \centering
    \includegraphics[width=1\linewidth]{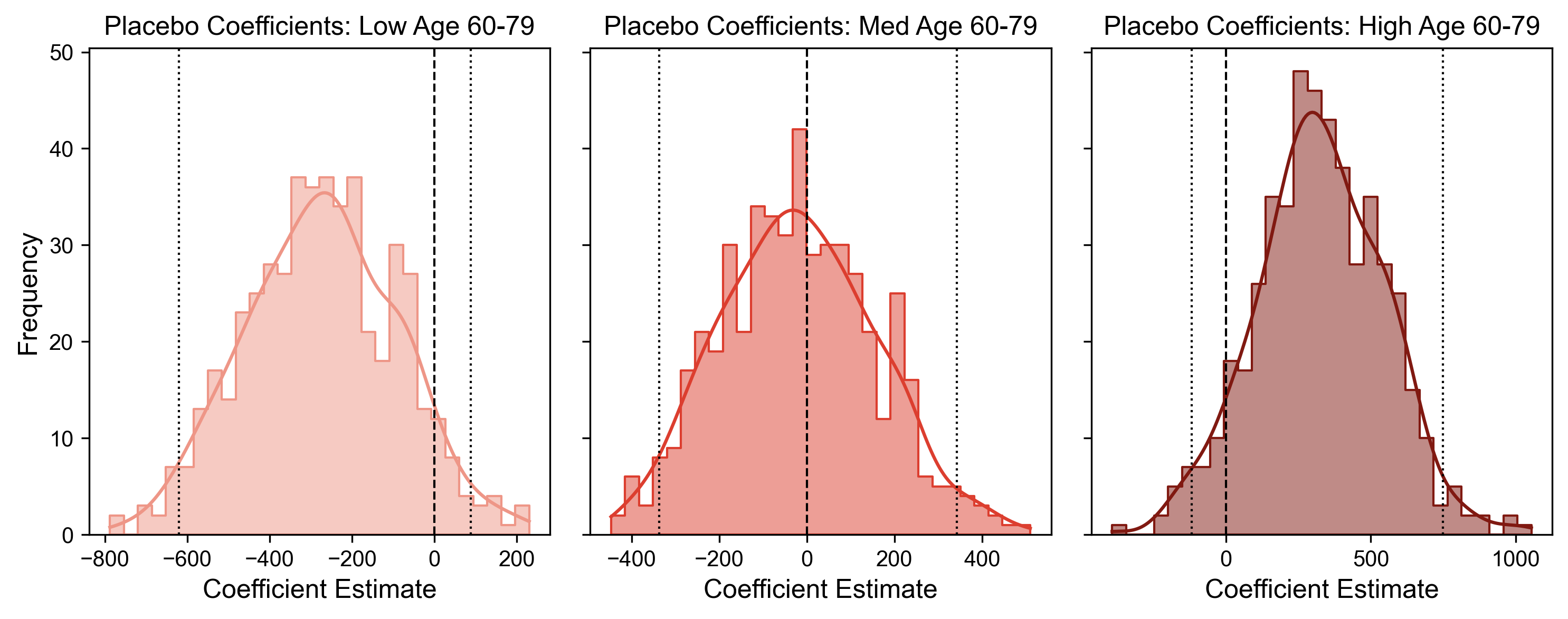}
    \caption[Placebo test coefficients for Age 60-79 \%]{\textbf{Placebo test coefficients for Age 60-79 \%.}}
    \label{fig:Figure S1}
\end{figure}

\begin{figure}[h]
    \centering
    \includegraphics[width=.65\linewidth]{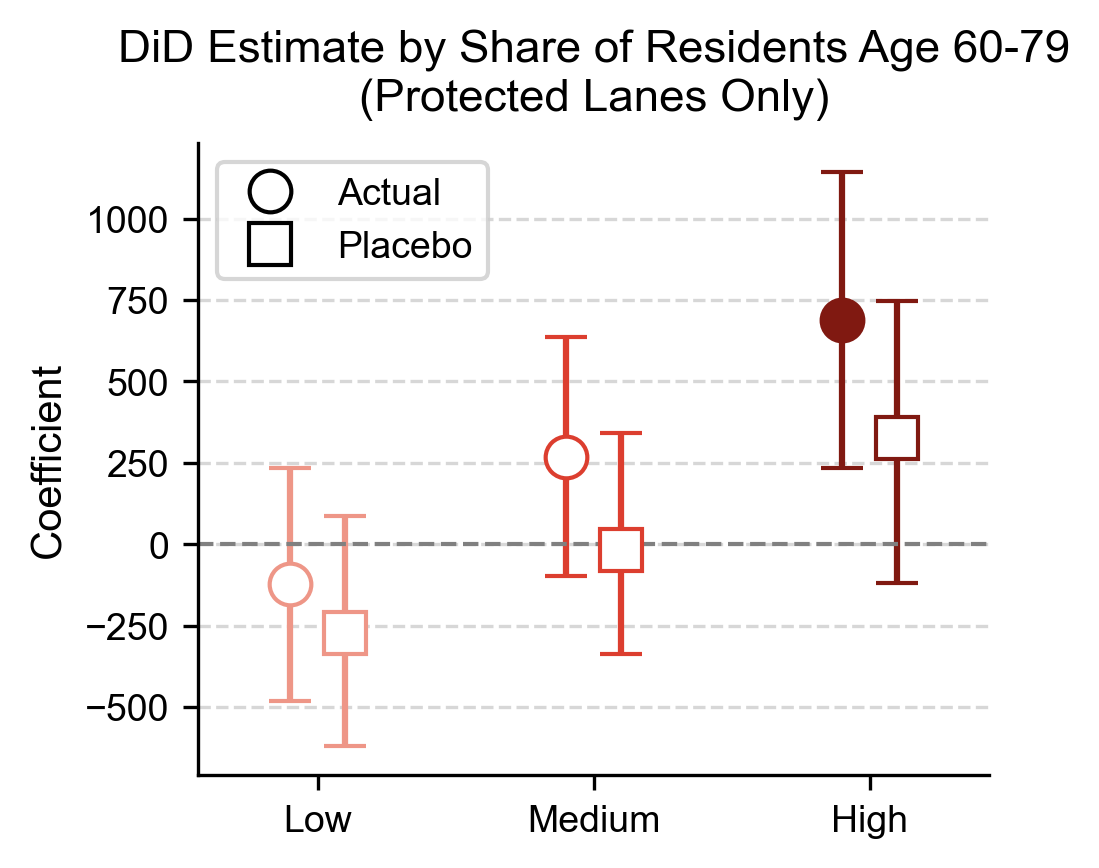}
    \caption[DiD results for Age 60-79 \%]{\textbf{DiD results for Age 60-79 \%.}}
    \label{fig:Figure S1}
\end{figure}

\begin{figure}[h]
    \centering
    \includegraphics[width=1\linewidth]{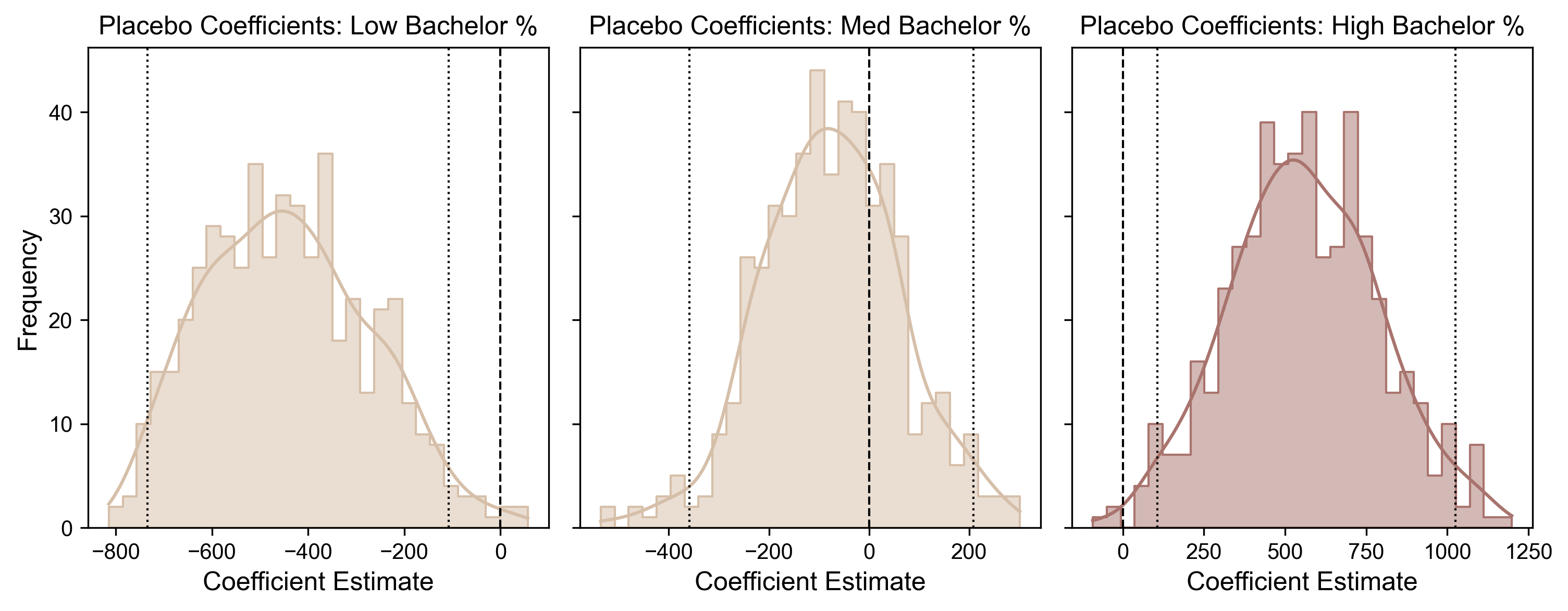}
    \caption[Placebo test coefficients for Bachelors \%]{\textbf{Placebo test coefficients for Bachelors \%.}}
    \label{fig:Figure S1}
\end{figure}

\begin{figure}[h]
    \centering
    \includegraphics[width=.65\linewidth]{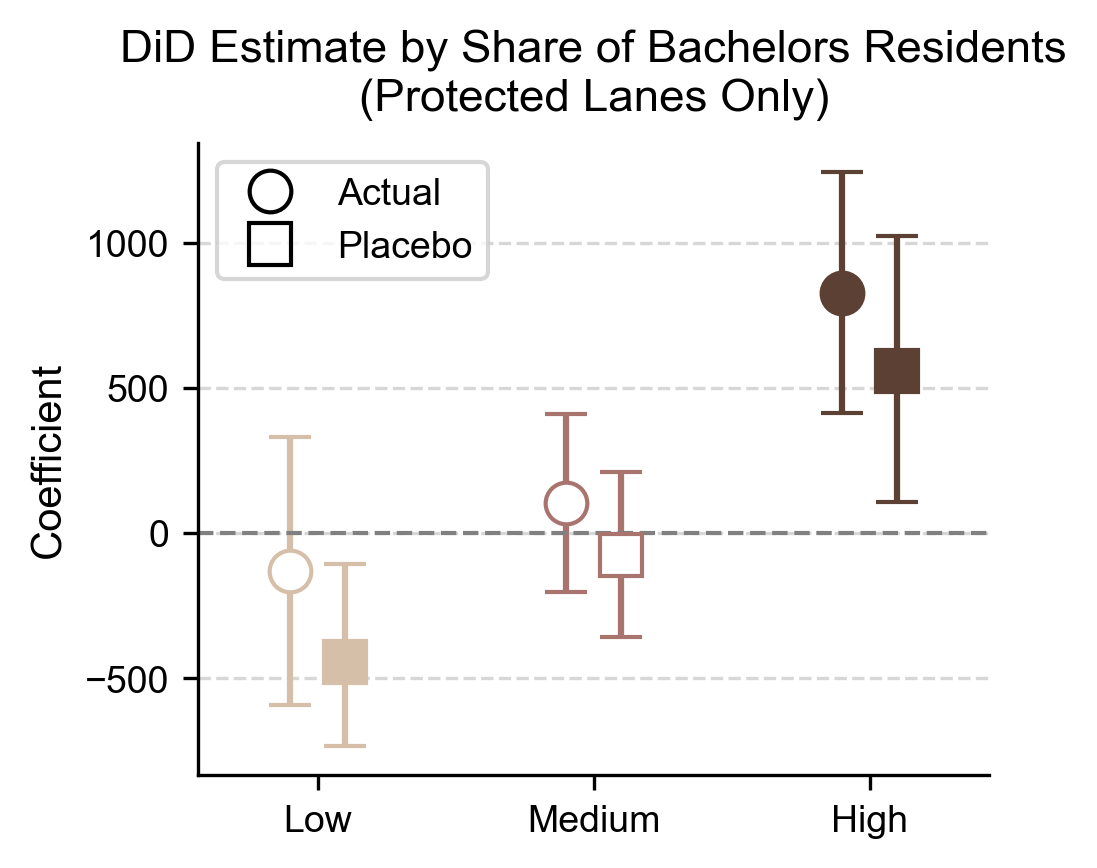}
    \caption[DiD results for Bachelors \%]{\textbf{DiD results for Bachelors \%.}}
    \label{fig:Figure S1}
\end{figure}
